\documentclass[11pt, a4paper]{article}
\pdfoutput=1 
\usepackage[height=8.85in,width=6.75in]{geometry}
\usepackage{graphicx,rotating}     %usepackage without driver option
\usepackage[bookmarksopen,colorlinks=true,linkcolor=teal,
citecolor=orangered,urlcolor=orangered,linktoc=all]{hyperref}

\usepackage{amsmath,amssymb}
\usepackage{slashed}
\usepackage{bm}
\usepackage{bbm}
\usepackage{cite}
\usepackage{soul}
\usepackage{comment}
\usepackage[utf8]{inputenc}
\usepackage{accents}
\usepackage[footnotesize]{caption}
\usepackage{cancel}
\usepackage{booktabs}
\usepackage{diagbox}
\usepackage{xstring}
\usepackage[most]{tcolorbox}

\makeatletter
\@addtoreset{equation}{section}

\makeatother

\usepackage{xcolor}
\definecolor{teal}{rgb}{0,0.502,0.502}
\definecolor{orangered}{rgb}{1,0.27,0.}
\definecolor{redorange}{rgb}{1,0.325,0.286}
\definecolor{forestgreen}{rgb}{0.13,0.55,0.13}

\usepackage{tikz}
\usepackage{tikz-feynman}
\tikzfeynmanset{compat=1.0.0}
\pgfdeclarelayer{bg}    % declare background layer
\pgfsetlayers{bg,main}  % set the order of the layers (main is the standard layer)

\newcommand{\IsBoldNumber}[1]{%
	\IfSubStr{1234}{#1}{\mathbf{#1}}{#1}%
}
\newcounter{i}
\newcommand{\bn}[1]{%
	\StrLen{#1}[\len]%
	\edef\result{}%
	\setcounter{i}{1}%
	\loop
	\StrChar{#1}{\value{i}}[\char]%
	\IsBoldNumber{\char}%
	\edef\result{\char}%
	\stepcounter{i}%
	\ifnum\value{i}<\numexpr\len+1
	\repeat
}
\begin{document}
%%%%%%%%%%%%%%%%%%%%%%%%%%%%%%%%%%%%%%%%%%%%%%%%%%%%%%%%%%%%
%%%%%%%%%%%%%%%%%%%%%%%%%%%%%%%%%%%%%%%%%%%%%%%%%%%%%%%%%%%%

%%%%%%%%%%%%%%%%%%%%%%%%%%%%%%%%%%%%%%%%%%%%%%%%%%%%%%%%%%%%
%%%%%%%%%%%%%%%%%%%%%%%%%%%%%%%%%%%%%%%%%%%%%%%%%%%%%%%%%%%%
\hypersetup{pageanchor=false}
\begin{titlepage}

 \begin{center}

 \hfill CERN-TH-2025-090\\
 \hfill UMN-TH-4423/25 \\
 \hfill FTPI-MINN-25-05\\

 \vskip 0.5in

 {\huge \bfseries On-shell recursion relations \vspace{5mm} \\ for higher-spin Compton amplitudes}%}
 \vskip .8in

 {\large Yohei Ema,$^{1,a}$} \let\thefootnote\relax\footnote{$^a$yohei.ema@cern.ch}
 {\large Ting Gao,$^{2,b}$} \footnote{$^b$gao00212@umn.edu}
 {\large Wenqi Ke,$^{2,3,c}$}
 \footnote{$^c$wke@umn.edu}
 {\large Zhen Liu,$^{2,d}$}
 \footnote{$^d$zliuphys@umn.edu}
 {\large Ishmam Mahbub$^{2,e}$}\footnote{$^e$mahbu008@umn.edu}
 \vskip .3in
 \begin{tabular}{ll}
 $^{1}$ & \!\!\!\!\!\emph{Theoretical Physics Department, CERN, 1211 Geneva 23, Switzerland}\\
 $^{2}$ & \!\!\!\!\!\emph{School of Physics and Astronomy, University of Minnesota, Minneapolis, MN 55455, USA}\\
 $^{3}$ & \!\!\!\!\!\emph{William I. Fine Theoretical Physics Institute, School of Physics and Astronomy,}\\[-.15em]
 & \!\!\!\!\!\emph{University of Minnesota, Minneapolis, MN 55455, USA}\\
 \end{tabular}

 \end{center}
 \vskip .6in

 \begin{abstract}
 \noindent
We recursively construct tree-level electromagnetic and gravitational Compton amplitudes of higher-spin massive particles 
by the all-line transverse momentum shift. 
With three-point amplitude as input, we demonstrate that higher-point electromagnetic and gravitational Compton amplitudes are on-shell constructible 
up to spin $s = 3/2$ and $s = 5/2$, respectively, under the all-line transverse shift after imposing the current constraint condition.
We unambiguously derive the four-point electromagnetic and gravitational Compton amplitudes for $s \leq 3/2$ and $s \leq 5/2$,
which are uniquely determined by the on-shell recursion relation and are free from unphysical spurious poles.  
In addition, we explore amplitudes of spin-$3/2$ particles with non-minimal three-point interactions with photon, as well as $s > 3/2$ particles, and comment on their notable features. Our work furthers the understanding of on-shell methods for massive amplitudes, with hopes to shed light on physical observables in particle physics and higher-spin amplitudes relevant for Kerr black-hole scattering.

  \end{abstract}

 \end{titlepage}
%%%%%%%%%%%%%%%%%%%%%%%%%%%%%%%%%%%%%%%%%%%%%%%%%%%%%%%%%%%%
%%%%%%%%%%%%%%%%%%%%%%%%%%%%%%%%%%%%%%%%%%%%%%%%%%%%%%%%%%%%

%%%%%%%%%%%%%%%%%%%%%%%%%%%%%%%%%%%%%%%
\tableofcontents
\renewcommand{\thepage}{\arabic{page}}
\renewcommand{\thefootnote}{$\natural$\arabic{footnote}}
\setcounter{footnote}{0}
%\newpage
\hypersetup{pageanchor=true}
%%%%%%%%%%%%%%%%%%%%%%%%%%%%%%%%%%%%%%%

%%%%%%%%%%%%%%%%%%%%%%%%%%%%%%%%%%%%%%%%%%%%%%%%%%%%%%%%%%%%
\section{Introduction }
%%%%%%%%%%%%%%%%%%%%%%%%%%%%%%%%%%%%%%%%%%%%%%%%%%%%%%%%%%%%
On-shell recursion relation exploits the analytic structure of the amplitude to construct higher-point amplitudes in terms of lower-point ones,
originally studied in the context of gauge theory and gravity via the Britto--Cachazo--Feng--Witten (BCFW) recursion relation~\cite{Britto:2004ap,Britto:2005fq}. 
Over the past decades, several recursion schemes have been developed and applied to broader classes of field theories~\cite{Bedford:2005yy,Cachazo:2005ca,Cohen:2010mi,Boels:2010bv,Cheung:2010vn,Arkani-Hamed:2010zjl,Gang:2010gy,Cheung:2015ota,Luo:2015tat,Elvang:2018dco,Guevara:2019ypd}. 
While significant progress has been made in massless theories, its extension to massive amplitudes remains an important challenge. 
Such an extension is crucial for studying the description of classical black hole dynamics~\cite{Levi:2015msa,Vines:2016qwa,Vines:2017hyw, Cachazo:2017jef,Guevara:2017csg,Arkani-Hamed:2017jhn,Guevara:2018wpp,Chung:2018kqs,Guevara:2019fsj,Johansson:2019dnu,Arkani-Hamed:2019ymq,Damgaard:2019lfh,Bern:2020buy,Aoude:2020onz,Bautista:2021wfy,Chiodaroli:2021eug,Chen:2021kxt,Cangemi:2022bew,Aoude:2022thd,Aoude:2022trd, Bern:2022kto, Bautista:2022wjf, Aoude:2023vdk, Cangemi:2023ysz,Cangemi:2023bpe,Bautista:2023sdf,Haddad:2023ylx, Chen:2023qzo, Bohnenblust:2024hkw}, amplitudes of Standard Model and Standard Model effective field theory (SMEFT)~\cite{Shadmi:2018xan,Christensen:2018zcq,Aoude:2019tzn,Durieux:2019eor,Christensen:2019mch,Durieux:2020gip,Balkin:2021dko,Liu:2023jbq,Christensen:2024bdt,Christensen:2024xzs}, 
massive supersymmetric amplitudes~\cite{Herderschee:2019dmc, Herderschee:2019ofc,Johansson:2023ymb} and massive gravity~\cite{Bonifacio:2018vzv,Bonifacio:2018aon,Bonifacio:2019mgk,Momeni:2020vvr,Gonzalez:2021ztm}, among others. 
Several recursive approaches have been explored for massive amplitudes, including massive BCFW recursion, all-line recursion techniques, and soft-recursion relations~\cite{Badger:2005zh,Badger:2005jv,Ozeren:2006ft,Boels:2007pj,Cohen:2010mi, Franken:2019wqr,Ballav:2020ese,Falkowski:2020aso,Wu:2021nmq,Ema:2024vww,Ema:2024rss,Gherghetta:2024tob}. 
In this paper, we apply the all-line transverse (ALT) shift, introduced in the context of QED and electroweak amplitudes~\cite{Ema:2024vww,Ema:2024rss} and then extended to supergravity~\cite{Gherghetta:2024tob}. We investigate on-shell construction of the tree-level electromagnetic and gravitational Compton amplitudes involving higher-spin (i.e. spin $s>1$) particles. 

As precision gravitational measurements become increasingly relevant, the theoretical description of higher-spin Compton amplitudes has gained recent attention, particularly in the study of gravitational wave scattering by Kerr black holes. The spinor-helicity formalism was introduced to describe these amplitudes in~\cite{Arkani-Hamed:2017jhn}. 
In the massive spinor-helicity formalism, a general on-shell three-point amplitude between photon/graviton and two equal-mass higher-spin particles can be constructed and further constrained by high-energy behavior. 
In higher-spin literature, it is well established that a consistent theory of massive higher-spin particles with a healthier 
high energy behavior requires the longitudinal current of the massive particle, $J^{\mu}$, 
to vanish in the massless limit: $\partial_{\mu}J^{\mu}=\mathcal{O}(m)$~\cite{Cucchieri:1994tx,Porrati:1993in}. 
This constraint selects the ``minimal'' amplitude, which exhibits less divergent UV behavior \cite{Chiodaroli:2021eug}. The resulting minimal amplitude is associated with a dipole moment $g=2$ and, in the high-energy limit, these amplitudes have the least divergent mass dependence~\cite{Chung:2018kqs}.

In the spinor-helicity formalism, four-point electromagnetic and gravitational Compton amplitudes
have been constructed from the minimal three-point amplitude through various approaches, including ``gluing'' three-point amplitudes 
(i.e. taking the products of three-point amplitudes without an explicit momentum shift)~\cite{Arkani-Hamed:2017jhn,Chung:2018kqs},\footnote{ Ref.~\cite{Chung:2018kqs} extends the discussion to non-minimal amplitudes to analyze constraints from consistent factorization.} BCFW recursion~\cite{Johansson:2019dnu}, and soft expansion~\cite{Guevara:2018wpp,Falkowski:2020aso}. 
For opposite helicity configurations of the external photons or gravitons, 
the four-point electromagnetic and gravitational Compton amplitudes constructed by gluing three-point amplitudes suffer from spurious pole contributions for $s>1$ and $s>2$, respectively, indicating corrections from contact terms. 
Similarly, the BCFW construction receives boundary contributions, which cannot be uniquely fixed without knowing the complete amplitude. 
To overcome these issues, the amplitude has been studied using Heavy Particle Effective Theory~\cite{Damgaard:2019lfh,Aoude:2020onz,Aoude:2022trd}. The contact term at the classical $\hbar \rightarrow 0$ limit is analyzed by requiring spin-shift symmetry in~\cite{Bern:2022kto,Aoude:2022trd}. 
Instead,~\cite{Chiodaroli:2021eug} determined the electromagnetic and gravitational amplitudes up to $s=3/2$
and $s = 5/2$, respectively, by imposing current constraints. 
 These ideas were then extended to construct a candidate four-point Compton amplitude for arbitrary spin~\cite{Cangemi:2023bpe,Bohnenblust:2024hkw}. The classical Compton amplitude has also been studied using general relativity calculations through black hole perturbation theory (BHPT)~\cite{Bautista:2021wfy,Bautista:2022wjf,Bautista:2023sdf}. 

In this paper, we study the on-shell construction of tree-level Compton amplitudes using the ALT shift. 
We analyze the large-$z$ behavior of these amplitudes through dimensional analysis. 
Since the shift deforms momentum via the transverse polarization vector, 
it exhibits improved large-$z$ behavior due to the Ward identity. This is crucial for the constructibility of the four-point scattering amplitudes for massive spin-1 and spin-3/2 particles~\cite{Ema:2024vww,Gherghetta:2024tob}. 
We demonstrate that the four-point electromagnetic and gravitational Compton amplitudes are on-shell constructible 
for $s \leq 3/2$ and $s \leq 5/2$ in the minimal case.\footnote{For amplitudes with $s > 3/2$ (photon coupling) or $s > 5/2$ (graviton coupling), and amplitudes with non-minimal couplings, the dimensional analysis does not guarantee on-shell constructibility.
} 
In our discussion, the current constraint, $\partial_{\mu} J^{\mu} = \mathcal{O}(m)$, plays a key role in the on-shell constructibility.
The external photons and gravitons also improve the large-$z$ behavior through their Ward identities,
which make higher-point amplitudes with multiple photons/gravitons on-shell constructible in the minimal case.
Due to the on-shell constructibility, we can uniquely fix the higher-point amplitudes by the lower-point amplitudes, 
and the final results are guaranteed to be free from spurious poles.
We demonstrate this explicitly by calculating the four-point amplitudes in the $s = 3/2$ photon and $s = 5/2$ graviton cases,
and our results match with those in~\cite{Chiodaroli:2021eug} obtained by a different method. We further extend our discussion to 
$s>3/2$ for photon and $s>5/2$ for graviton Compton amplitudes, and show that these amplitudes are not constructible under the ALT shift,
with the minimal three-point amplitudes as the input.
Although we focus only on the four-point Compton amplitudes in this paper, a virtue of the on-shell recursive method is that it is
straightforward to extend the calculation to higher-point amplitudes.
Indeed, our dimensional analysis indicates that higher-point Compton amplitudes, with $n > 2$ external photons or gravitons,
are also constructible for $s \leq 3/2$ and $s \leq 5/2$ for the electromagnetic and gravitational cases, respectively,
regardless of the helicity combinations of the external photons or gravitons.
Therefore, we expect that higher-point Compton amplitudes for mixed helicity cases are constructible
without spurious poles and contact term ambiguities by the ALT shift, 
which potentially improves our understanding of the higher-spin scattering amplitudes.
This paper can thus be understood as an important step towards the on-shell recursive construction of higher-point 
higher-spin Compton amplitudes.

Our dimensional analysis does not establish the on-shell constructibility for non-minimal three-point amplitudes, 
as these do not satisfy the current constraint.
Nevertheless, we find a set of the factorized part of four-point Compton amplitudes constructed from the non-minimal three-point amplitudes 
by the ALT shift, independent of the shifted momentum and hence little-group covariant by themselves.
These findings might suggest a larger set of on-shell constructible amplitudes beyond the minimal one.

This paper is organized as follows. In Sec.~\ref{sec:overview}, we review the spinor-helicity formalism and provide the three-point amplitudes from which we construct the higher-point amplitudes. We define the momentum shift used throughout this paper, the ALT shift, at the end of this section.
In Sec.~\ref{sec:photoncompton}, we establish the on-shell constructibility of the four-point electromagnetic Compton amplitudes 
for $s \leq 3/2$ by the dimensional analysis.
We explicitly calculate the four-point amplitude for $s = 3/2$, and obtain a little-group covariant result free from spurious poles.
Although we establish the constructibility only for the minimal amplitudes up to $s = 3/2$,
we comment on the case of $s > 3/2$, in particular $s = 2$. 
In Sec.~\ref{sec:comptongraviton}, we establish the on-shell constructibility 
and calculate the four-point gravitational Compton amplitudes for $s = 5/2$ by the ALT shift,
in a parallel manner to the electromagnetic case.
Finally, Sec.~\ref{sec:conclusion} is devoted to the conclusion and discussion.  
Additional information can be found in the appendices.
We summarize our conventions and review the spinor-helicity formalism in App.~\ref{app:convention}. 
App.~\ref{app:threepoint} provides the spin-3/2 photon and -5/2 graviton three-point amplitudes 
used in the paper. We make a connection with possible Lagrangian interactions there.

%%%%%%%%%%%%%%%%%%%%%%%%
\section{Overview of the  formalism }\label{sec:overview}
\subsection{Spinor helicity formalism}
%%%%%%%%%%%%%%%%%%%%%%%%

In this subsection, we review the little-group covariant construction of amplitudes using the \( SL(2, \mathbb{C}) \) spinors. 
We map four-momentum $p^{\mu}$ as
\begin{align}
	p_{\alpha\dot{\alpha}} = p_\mu \sigma^\mu_{\alpha \dot{\alpha}} = 
	\begin{pmatrix} p^0 - p^3 & -p^1 + i p^2 \\ -p^1 - ip^2 & p^0 + p^3\end{pmatrix} \ ,
 \label{eq:spinor_momentum}
\end{align}
where we take the metric as $	\eta_{\mu\nu} = \mathrm{diag}(+1, -1, -1, -1)$, and $\alpha, \dot{\alpha}=1,2$ are $SL(2,\mathbb{C})$ indices. 
For massless particle, $\mathrm{det}\,p_{\alpha\dot{\alpha}} = 0$, so $p_{\alpha\dot{\alpha}} $ can be written as a product of two spinors
%%%%%%%%%%%%%%
\begin{equation}
	p_{\alpha\dot{\alpha}} \ = \ \lambda_\alpha \tilde{\lambda}_{\dot{\alpha}} \ ,
        \label{eq:spinor}
\end{equation}
%%%%%%%%%%%%%
where $ \lambda_{\alpha} $ and $\tilde{\lambda}_{\dot{\alpha}}$ transform as the $(1/2,0)$ and $(0,1/2)$  representations of $SL(2,\mathbb{C})$, respectively.  If the momentum is real, $p_{\alpha\dot{\alpha}}^{\dagger}=p_{\alpha\dot{\alpha}}$, 
the spinors are related as $\lambda_{\alpha}^*  = \pm \tilde{\lambda}_{\dot{\alpha}}$. 
The spinor description has redundancy, corresponding to the little group freedom which
leaves the momentum invariant. Indeed,
we can rescale $\lambda_a, \tilde{\lambda}_{\dot{a}}$ by a complex phase $w$ as
%%%%%%%%%%
\begin{equation}
    \lambda_a \rightarrow w^{-1}   \lambda_a \ \ \ \tilde{\lambda}_{\dot{a}} \rightarrow w \tilde{\lambda}_{\dot{a}},
    \label{eq:little_group}
\end{equation}
%%%%%%%%%%%%
corresponding to the $U(1)$ little group. For massive momentum $\mathrm{det}\,p_{\alpha\dot{\alpha}}=m^2 \neq 0$ and so we can write $p_{\alpha \dot{\alpha}}$ as a summation of two rank-one matrices, each of which is written as a product of two spinors
\begin{align}
	p_{\alpha \dot{\alpha}} = \lambda_{\alpha}^I \tilde{\lambda}_{\dot{\alpha}I},
	\quad
	I = 1,2  .
\end{align}
We contract indices using the completely antisymmetric tensor $\varepsilon^{IJ}$ with $\varepsilon^{12}=1$.  Momentum is invariant under the little group transformation
%%%%%%%%%%%
\begin{align}
     \lambda_{\alpha}^I  \rightarrow {W^I}_J  \lambda_{\alpha}^J, \quad   \tilde{\lambda}_{\dot{\alpha}}^I  \rightarrow {W^I}_J  \tilde{\lambda}_{\dot{\alpha}}^J ,
\end{align}
%%%%
where $W \in SU(2)$. We use the bold notation to represent massive spinors following~\cite{Arkani-Hamed:2017jhn}: $\lambda_{i\alpha}^I = \vert \textbf{i}\rangle_{\alpha}^I$ and $\tilde{\lambda}_{i\dot{\alpha}}^I = [\textbf{i}\vert_{\dot{\alpha}}^I$ for spinors correspond to a momentum $p_i$ of a particle $i$. Unless explicitly stated, we suppress the little-group index.  The spinors satisfy
\begin{align}
 (p_i)^{\dot{\alpha}\alpha}{\lambda}^{I}_{i\alpha}=   {p}_i \lvert \textbf{i} \rangle^I = m_i \lvert \textbf{i}]^I&, 
    \quad 
   (p_i)_{\alpha\dot{\alpha}}\tilde{\lambda}^{\dot{\alpha}I} ={p}_i \lvert \textbf{i} ]^I = m_i \lvert \textbf{i}\rangle^I.
\end{align}
These equations relate $\lambda_i$ and $\tilde{\lambda}_i$, and hence amplitudes are expressible with solely $\lambda_i$ or solely $\tilde{\lambda}_i$. One can construct a spin-$s$ representation using $2s$ spinors, $\lambda_i$ and $\tilde{\lambda}_i$, by symmetrizing their little group indices. 
Factoring out the external $\lambda_i$ and $\tilde{\lambda}_i$ spinors of a particle $i$, we get
%%%%%%
\begin{align}
    \mathcal{A}_n^{...\{I_{i_1},I_{i_2},...,I_{i_{2s}}\}...} = ... \lambda_{i_1}^{\{I_{i_1}} \tilde{\lambda}_{\dot{i}_2}^{I_{i_2}} ... \lambda_{i_{2s}}^{I_{i_{2s}}\}} ...{A}^{...i_1,\dot{i}_2,...,i_{2s},..}_n,
\end{align}
%%%%%%%%%%%%$
where the most external ``$\cdots$'' indicates the little-group indices and the external spinors of particles other than the particle $i$,
and ``$\{\cdots\}$'' indicates the symmetrization with respect to the indices.
This form makes the little-group covariance of the amplitude manifest.

%%%%%%%%%%%%%%%%%%%%%%%%%%%%%%%%%%%%%%%%%%%%%%%%%%%%%%
 \subsection{Three-point amplitudes with one massless particle}
%%%%%%%%%%%%%%%%%%%%%%%%%%%%%%%%%%%%%%%%%%%%%%%%%%%%%%
The general form of the on-shell three-point amplitude ${A}_3(\psi^s_1, \Bar{\psi}^s_2, A_3^h)$
involving one massless particle $A_3$ with helicity $h$ and two equal-mass particles $\psi^s_1$ and $\bar{\psi}_2^s$, each with spin $s$, 
is discussed in~\cite{Arkani-Hamed:2017jhn, Chung:2018kqs}. To express the amplitude, we first note that the spinors representing the massless particle, 
$\lvert 3 \rangle_{\alpha}$ and $[3\lvert_{\dot{\alpha}}$, are linearly dependent since
%%%%
\begin{align}
    p_2^2 -m^2 = 2 p_1 \cdot p_3 = - 2 p_2 \cdot p_3 = [ 3 \lvert p_1 \lvert 3 \rangle = - [ 3 \lvert p_2 \lvert 3 \rangle = 0 \ .
\end{align}
%%%%%%%%%%%
We therefore introduce a dimensionless proportionality constant $x_{12}$ as
%%%%%%%%%%%%%%
\begin{align}
    x_{12}\lvert 3 \rangle &= \frac{p_1-p_2}{2m }|3] , \quad
    \tilde{x}_{12} \lvert 3 ] = \frac{p_1-p_2}{2m}|3\rangle \ ,
\end{align}
%%%%%%%%%%%%
where $m$ denotes the mass of $\psi^s_1$ and $\bar{\psi}_2^s$. Note that $x_{12}$ has helicity weight $+1$ and $\tilde{x}_{12} = 1/x_{12}$
has helicity weight $-1$.
General amplitudes depend on the massless polarization vector through these $x$-factors. We can then write the three-point amplitude as\footnote{
See~\cite{Guevara:2018wpp,Chung:2018kqs} for the three-point amplitude in the (anti-)chiral basis and its connection to the multipole expansion.}
%%%%%%%%%%%%%%%%%
\begin{align}
{A}_3(\psi_1^{s},\bar{\psi}_2^{s},A^{-h}_3) =\frac{\tilde{x}^h_{12} }{m^{2s-1}}\bigg( {c}_1  [\textbf{1} \textbf{2} ]^{2s} + {c}_2 [\textbf{1} \textbf{2} ]^{2s-1}\langle \textbf{1}\textbf{2}\rangle + ... +{c}_{2s}  \langle \textbf{1}\textbf{2}\rangle^{2s} \bigg) \ .
\label{eq:mult_exp3}
\end{align}
%%%%%%%%%%%%%%%%
We may fix the coefficients $c_i$ from high-energy behavior of the amplitude. The minimal amplitudes\footnote{To avoid the confusion of terminologies, in our work,  the word ``minimal'' coupling or amplitude refers to the terms~\eqref{eq:min}, whereas in higher spin literature, ``minimal coupling'' is generally defined/obtained via the replacement $\partial\rightarrow \mathcal{D}$. The resulting interactions differ from our definition for $s>1$.}
%%%%%%%%
\begin{align}
    {A}_3(\psi_1^{s},\bar{\psi}_2^{s},A^{-h}_3) &=\tilde{c}\,\tilde{x}^h_{12}   \frac{[\textbf{1} \textbf{2} ]^{2s}}{m^{2s-1}}\ ,
    \quad
    {A}_3(\psi_1^{s},\bar{\psi}_2^{s},A^{h}_3) =c\,{x}^h_{12}   \frac{\langle \textbf{1} \textbf{2} \rangle^{2s}}{m^{2s-1}}\ , 
    \label{eq:min}
\end{align}
%%%%%%%%%
corresponding to setting $c_i = 0$ for $i \geq 2$ or $i \leq 2s-1$, 
have improved UV behavior and correspond to dipole moment $g=2$~\cite{Chung:2018kqs}. It is suggested that these minimal amplitudes for photon and graviton coupling arise from higher-spin theories that satisfy the off-shell current constraint $P \cdot J = \mathcal{O}(m)$ for arbitrary spin~\cite{Chiodaroli:2021eug}. The correspondence between the minimal amplitude and the current constraint can be verified up to spin-3/2 gauge theory and spin-5/2 gravity.

\subsection{On-shell recursion relation}
On-shell recursion relation constructs higher-point amplitudes in terms of lower-point on-shell amplitudes. 
The central idea is to shift external momenta systematically by a complex parameter $z$, %
%%%%%%
\begin{equation}
    \hat{p}_i (z)  = p_i + z  q_i\,.
 \label{eq2}
\end{equation}
%%%%%%
We demand that ${\hat{p}_i(z)}$ keeps the on-shell condition intact and satisfies the total momentum conservation,
\begin{align}
	\hat{p}_i^2(z) = p_i^2 = m_i^2,
	\quad
	\sum_i \hat{p}_i(z) = 0,
\end{align}
which restricts the choice of $q_i$. 
The momentum shift transforms an $n$-point amplitude to a complex function of $z$, $\hat{A}_n(z)$. 
Using Cauchy's theorem, we find
\begin{align}
\label{Cauchy theorem}
	 {A}_n &= \frac{1}{2\pi i}\oint_{z=0} \frac{dz}{z} \hat{A}_n(z)
  = - \sum_{\{z_I\}} \text{Res}\left[\dfrac{\hat{A}_n(z)}{z}\right] + B_\infty \ ,
\end{align}
where the integration contour is deformed to infinity in the second equality, picking up the poles outside $z=0$, which includes poles $\{z_I\}$ at finite $z$ and a possible boundary contribution $B_\infty$ at $\lvert z \lvert \rightarrow \infty$. 
The pole corresponds to an intermediate particle going on-shell, $\hat{p}_I^2 (z_I) = m_I^2$, where the amplitude factorizes into lower-point on-shell sub-amplitudes, leading to
\begin{align}
	A_n = -\sum_{z = z_I}\sum_s\mathrm{Res}\left[\hat{A}^{(s)}_{n-m+2}\frac{1}{z}\frac{1}{\hat{p}_I^2 - m_I^2}
	\hat{A}^{(-s)}_{m}\right]
	+ B_\infty.
	\label{eq:recursion_general}
\end{align}
The factorized part satisfies $3 \leq m \leq n-1$, and $s$ is the projection of spin along the choice of spin axis. The amplitude can then be constructed solely from lower-point on-shell amplitudes if 
%%%%%%%
\begin{align}
	B_\infty = 0, ~~\mathrm{or}~~ \hat{A}(z) \to 0~~\mathrm{at}~~z \to \infty \ .
	\label{eq:constructibility}
\end{align}
%%%%%
Both the underlying theory (specific models) and the choice of $\{q_i\}$ (specific momentum shift schemes) determine if $B_\infty = 0$. 
If $B_\infty = 0$ with a given shift scheme, a theory is \textit{on-shell constructible} under this shift. From a Lagrangian-based viewpoint, $B_\infty = 0$ implies that any contact term present in the Lagrangian arises from the factorized part of the on-shell amplitude in the recursion method. However, \textit{independent} contact terms can exist that are not uniquely fixed by the information and consistency of lower-point interactions.

%%%%%%%%%%%%%%%%%%%%%%%%%%%%%%%%%%%%%%%%
\subsection{All-line transverse  shift}
\label{sec:alt_shift}
%%%%%%%%%%%%%%%%%%%%%%%%%%%%%%%%%%%%%%%%
Throughout our paper, we stick to the ALT momentum shift~\cite{Ema:2024vww,Ema:2024rss}. 
An important feature of the ALT shift is a well-defined large-$z$ behavior of the amplitude, crucial to see if $B_\infty = 0$,
controlled by the external particle species and the dimensionality of the interactions (see Section~\ref{sec:constructibility}).
Under the ALT shift, all external momenta are shifted by their transverse polarization vectors.   

The shift for general helicity amplitudes is most conveniently defined in the helicity basis, where we write the massive spinors as
\begin{align}
   \lvert \textbf{i} \rangle^I_a = \lvert i \rangle_a \delta^I_- + \lvert \eta_i \rangle_a \delta^{I}_+, \quad [ \textbf{i} \lvert^I_{\dot{a}} = [ i \lvert_{\dot{a}} \delta_+^I + [\eta_i \lvert _{\dot{a}} \delta^{I}_-,
\end{align}
so that the little group indices $I = 1$ and $I=2$ correspond to the $+$ and $-$ helicity states, respectively. In this choice of basis, the spinors $\eta$, $\tilde{\eta}$ scale as $m$ in the small mass limit  (see Appendix \ref{app:convention} for more details). Now, the ALT shift is defined for the transverse modes of a massive spin-$s$ particle as

\begin{align}
	\begin{cases}
	\vert i \rangle \to \vert \hat{i}\rangle = \vert i\rangle + z c_i \vert \eta_i\rangle
	& \mathrm{for}~~\lambda_i = +s, \vspace{1mm} \\
	\vert i ] \to \vert \hat{i}] = \vert i] + z c_i \vert \eta_i]
	& \mathrm{for}~~\lambda_i = -s,
	\end{cases}
      \quad \mathrm{or}~~~
	\begin{cases}
	\vert \eta_i ] \to \vert \hat{\eta}_i] = \vert \eta_i] - z c_i \vert i]
	& \mathrm{for}~~\lambda_i = +s, \vspace{1mm} \\
	\vert \eta_i \rangle \to \vert \hat{\eta}_i\rangle = \vert \eta_i\rangle - z c_i \vert i\rangle
	& \mathrm{for}~~\lambda_i = -s,
	\end{cases}	
 \label{transverseshifts}
\end{align}
and the other spinors are unshifted. Here $\lambda_i$ is the helicity of the particle $i$, and $c_i$ are constants, constrained
to maintain the total momentum conservation. 
The momentum of a massive particle is expressed in the helicity basis as
\begin{align}
    (p_i)_{\dot{a}a}=\vert i\rangle_{a}[ i\vert_{\dot{a}}-\vert \eta_i\rangle_{a}[\eta_i\vert_{\dot{a}}.
\end{align}
The momentum is therefore shifted by $zc_i\vert \eta_i\rangle[ i\vert$ for $\lambda_i = +s$ and $zc_i\vert i\rangle[ \eta_i\vert$ for $\lambda_i = -s$, 
which are proportional to the transverse polarization vector of the corresponding helicity state. The on-shell condition of~$\hat{p}_i$, $\hat{p}_i^2 = p_i^2 = m_i^2$, is then satisfied automatically.
The transverse spinors and polarization vectors of external massive particles 
are not shifted by the ALT shift, which simplifies the large-$z$ behavior of the amplitudes,
and therefore we focus on the scattering of transverse modes in the following.
The result can be generalized to the amplitudes with general helicity configurations thanks to the little group covariance of the amplitude,
by acting the spin-raising and spin-lowering operators
\begin{align}
	     J_i^- = -\lvert i \rangle_a \dfrac{\partial}{\partial \lvert \eta_i \rangle_a} 
	     - [ \eta_i \lvert_{\dot{a}} \dfrac{\partial}{\partial [ i\lvert_{\dot{a}}}, 
	     \qquad
	     J_i^+ = \lvert \eta_i \rangle_a \dfrac{\partial}{\partial \lvert i \rangle_a} 
	     + [ i \lvert_{\dot{a}} \dfrac{\partial}{\partial [ \eta_i\lvert_{\dot{a}}},
         \label{eq:spin-raising}
\end{align}
on the transverse amplitudes\footnote{ Note that for the remaining $2s-1$ spin/helicity states, the shift would introduce additional $z$-dependence on their polarization vectors. Such an additional $z$-dependence will complicate the constructibility analysis in later sections. Although there are various possibilities to refine the shift for different states, we find that the most convenient approach is to use raising and lowering operators once the $\lambda_i=\pm s$ amplitudes are obtained.
}.

For massless particles, 
one can choose reference spinors $\vert \xi_i ]$ and $\vert \xi_i \rangle$ to perform the ALT shift:
\begin{align}
	\begin{cases}
	\vert i \rangle \to \vert \hat{i}\rangle = \vert i\rangle + z c_i \vert \xi_i\rangle
	& \mathrm{for}~~\lambda_i = +s, \vspace{1mm} \\
	\vert i ] \to \vert \hat{i}] = \vert i] + z c_i \vert \xi_i]
	& \mathrm{for}~~\lambda_i = -s.
	\end{cases}
\end{align}
For massless spinors, the polarization vector is fixed only up to a reference spinor $\zeta_i$ as
\begin{align}
	\epsilon_i^{(+)} = \sqrt{2}\frac{\vert \zeta_i\rangle [ i \vert}{\langle i \zeta_i\rangle},
	\quad
	\epsilon_i^{(-)} = -\sqrt{2}\frac{\vert i\rangle [ \zeta_i \vert}{[i\zeta_i]}.
\label{polamassless}\end{align}
The momentum, on the other hand, is shifted by $\xi_i$, which could be different from $\zeta_i$:
\begin{align}
    (p_i)_{\dot{a}a}=\vert i\rangle_{a}[ i\vert_{\dot{a}}\quad \to \quad
    \begin{cases}
        \hat{p}_i=p_i+zc_i\vert \xi_i\rangle[ i\vert & \mathrm{for}~~\lambda_i = +s, \vspace{1mm} \\
        \hat{p}_i=p_i+zc_i\vert i\rangle[ \xi_i\vert & \mathrm{for}~~\lambda_i = -s.
    \end{cases}
\end{align}
If one chooses $\xi_i = \zeta_i$, the polarization vector~\eqref{polamassless} is again invariant under the ALT shift.
We can instead choose $\xi_i \neq \zeta_i$, which shifts the denominator of the polarization vectors 
and improves the large-$z$ behavior of the amplitude by $1/z$ for each external massless polarization vector.
This does not give rise to a spurious pole as a different choice of the reference vector is 
a gauge artifact and the full result is independent of this choice.
This improved large-$z$ behavior plays an important role in the constructibility of the Compton amplitude.\footnote{
	For the former choice, the improved large-$z$ behavior 
	follows from that the momentum becomes proportional to the polarization vector in the large-$z$ limit. 
	The amplitude then vanishes in the large-$z$ limit at the leading order 
	due to the Ward identity, indicating the improved large-$z$ behavior~\cite{Ema:2024rss}. 
	Note that $\lim_{z\to\infty} \hat{p}_i \propto \epsilon_i^{(\pm)}$ only if $\xi_i = \zeta_i$.
}

The total momentum conservation puts four constraints on $c_i$. 
For massive external particles, the transverse polarizations in the helicity basis do not have a temporal component, and therefore only three constraints are independent if all external particles are massive. For massless external particles, the choice of $\xi_i$ gives additional degrees of freedom in satisfying the momentum conservation. These two observations together guarantee the existence of non-trivial solutions for four- and higher-point amplitudes.
Eq.~\eqref{eq:recursion_general} involves the shifted spinors and momenta. It can be evaluated either directly by solving for $c_i$ explicitly, or indirectly by simplifying the expression to a form such that $c_i$ cancels out, without using the explicit solutions of $c_i$. We take the latter approach in this paper that makes several important properties of the final amplitudes transparent, while it is numerically easier to implement the former approach for general amplitude constructions.

The ALT shift breaks the little group covariance of the amplitudes at the intermediate steps since $q_i \propto \epsilon_i^{(\pm)}$ depends on the little group indices.
If $B_\infty = 0$, this dependence must cancel out in the final result due to the little group covariance of the original amplitudes.
Therefore, any residual $q_i$ dependence of the factorized part (i.e. the first term in Eq.~\eqref{eq:recursion_general}) means $B_\infty \neq 0$.
Note that this is only a necessary condition and not a sufficient condition; the cancellation of the $q_i$ dependence 
does not guarantee $B_\infty = 0$ in general, 
but it indicates the potential existence of a theory that realizes the resultant amplitude with $B_\infty = 0$, and hence is constructible.

%%%%%%%%%%%%%%%%%%%%%%%%%%%%%%%%%%%%%%%
\section{Photon Compton amplitude}
\label{sec:photoncompton}
%%%%%%%%%%%%%%%%%%%%%%%%%%%%%%%%%%%%%%%

In this section, we construct the photon Compton amplitudes for the transverse massive spin-$s$ particles, with $s \leq 3/2$, using the ALT shift. 
Our result is little-group covariant, and therefore, we can utilize the spin-raising and spin-lowering operators in Eq.~\eqref{eq:spin-raising} to obtain the amplitude with longitudinal modes after deriving the amplitude for transverse modes. 

%%%%%%%%%%%%%%%%%%%%%%%%%
%%%%%%%%%%%%%%%%%%%
\subsection{Constructibility of Compton amplitudes}
\label{sec:constructibility}

To construct the amplitudes recursively, it is crucial to show that $B_\infty = 0$.
Therefore, we first examine the large-$z$ behavior of the Compton amplitude using dimensional analysis. We decompose the $n$-point amplitude into a kinematic part and external wavefunctions/polarization vectors as~\cite{Cheung:2015cba, Ema:2024vww, Ema:2024rss}
%%%%%%%%%%%%%%%%%%
%%
\begin{align}
	{A}_n = \left(\sum_{\mathrm{diagrams}}g \times F\right) \times \prod_{\mathrm{vector}} \epsilon \times \prod_{\mathrm{spinor}}u \ ,\label{generalamp}
\end{align}
where $g$ represents a product of coupling constants and $F$ represents the kinematic factors. 
Since the polarization vector of a higher-spin particle can be represented using spin-1 polarizations $\epsilon$ and Dirac spinors $u$ (in the case of a half-integer spin), we collectively express them as $\prod \epsilon \times \prod u$
to identify the mass dimensionality counting of the resulting amplitudes from the Dirac spinors. Later, we will represent the amplitudes by the symmetrized product of massive spinors when performing the shift.

We focus on the large-$z$ behavior of the $\pm s$ helicity (i.e. transverse) amplitude of the spin-$s$ particle. 
For these modes, the external massive polarization vectors and spinors are not altered by the ALT shift, 
so their contribution to $\prod \epsilon \times \prod u$ does not affect the large-$z$ behavior. 
The kinematic factor $F$ can be expressed as $F = N/D$, where $D = \prod_I (\hat{p}_I^2 - m_I^2) $ 
with $I$ indicating the intermediate particles. For the ALT shift, the denominator of each propagator introduces a factor of $z^2$. 
Then, the $z$ scaling of $D$ is given by
\begin{align}
    \lim_{|z|\rightarrow \infty} \hat{D} \sim z^{[D]} \  ,
\end{align}
where $[D]$ is the mass dimension of $D$. All the additional kinematic components from the propagator's numerators and/or derivative couplings are contained in $N$. 
Suppose that the total number of momentum insertions in the numerator does not exceed the mass dimension of $N$.
Then, we have 
%%%%%%%%%%%%%%%
\begin{align}
        \lim_{|z|\rightarrow \infty} \hat{N} \sim z^{\gamma_N}, \quad  \gamma_N \leq [N] \ ,
\end{align}
%%%%%%%%%%%%%%%
so that $F$ scales as $\hat{F} \sim z^{\gamma}$ with $\gamma \leq [F]$. 
The mass dimension of ${A}_n$, $4-n$, determines the mass dimension of $F$ as
%%%%%%%%%%%%%%%%%%
\begin{align}
    [F] = 4-n - [g] - \frac{N_F}{2} \ ,
\end{align}
%%%%%%%%%%%%%%%%%
where $N_F$ denotes the total number of external spinors. Thus, the $z$ scaling of the $n$-point amplitude is bounded as
\begin{align}
 \lim_{|z|\rightarrow \infty} \hat{{A}}_n \sim z^{\gamma}, \quad \gamma \leq  4-n - [g] - \frac{N_F}{2} \ .
 \label{eq:large-z-dim}
\end{align}
%%%%%%%%%%%%%%%%%
Note that, the dimension of $[g]$ accounts for the $1/\Lambda$ divergences and $z$-scaling from higher-dimensional operators. The assumption that the total number of momentum insertions in the numerator does not exceed the mass dimension of $N$ 
breaks down if \( F \) exhibits any mass divergence as \( m \to 0 \).
This may occur in the presence of non-minimal interactions when the current constraint \( \partial \cdot J = \mathcal{O}(m) \) is not satisfied. In such cases, additional terms of the form $\hat{p}_I/m_I$ arise from the massive propagators, potentially leading to a $1/m$-dependence in \( F \). Therefore, our analysis is not directly applicable to theories that do not satisfy current constraint. 

In the case of the Compton amplitudes, the external photons and gravitons further improve the large-$z$ behavior 
through their external polarization vectors/tensors as argued in Sec.~\ref{sec:alt_shift}. 
If we set $n=4$, the large-$z$ behavior for the electromagnetic Compton amplitudes is
\begin{equation}
    \gamma\leq -[g]-\frac{N_F}{2}-2,
\label{photonscaling}\end{equation}
and that for the gravitational Compton amplitudes is, 
\begin{equation}
    \gamma\leq -[g]-\frac{N_F}{2}-4.
\end{equation}
An amplitude is constructible if $\gamma <0$. 
Therefore, the four-point Compton amplitudes are constructible for any renormalizable theories with $[g]\geq 0$, such as electron and $W$ boson Compton scatterings. For the spin-$3/2$ electromagnetic Compton scattering, the couplings are non-renormalizable, $[g] \leq -2$, but as long as (i) the three-point amplitude is the minimal one satisfying the current constraint $\partial\cdot J=\mathcal{O}(m)$, 
and (ii) there is no more than one momentum insertion in the interaction, we can use Eq.~\eqref{photonscaling} to conclude that $\gamma\leq -1$ and hence such an amplitude is constructible under the ALT shift. 
Similarly, for the minimal three-point amplitude with $s \leq 3/2$, all the higher-point Compton amplitudes with multiple photons are constructible, as the inclusion of additional massless polarizations and propagators further improves the large-$z$ behavior. 

The above argument tells us that increasing the dimension of the operators, or $-[g]$, hinders the on-shell constructibility. 
This is expected since higher-dimensional operators correspond to independent higher-point contact terms. 
For particles with $s>3/2$, one expects such higher-dimensional operators to be present, but their precise forms are unknown due to the lack of knowledge about charged higher-spin Lagrangians. 
We will see that Compton amplitudes are not constructible for $s \geq 2$ in the minimal case under the ALT shift since the factorized part has the residual $q_i$-dependence.
We note that our analysis provides only an \textit{upper bound} on the $z$-scaling of the amplitude; the actual large-$z$ behavior of the amplitude could be better.

%%%%%%%%%%%%%%%%%%%
\subsection{Four-point Compton amplitude}
%%%%%%%%%%%%%%%%%%%

We now recursively construct the four-point Compton amplitude, with the minimal three-point amplitudes Eq.~\eqref{eq:min} as an input
(later we extend it to non-minimal three-point amplitudes).
We label the amplitude as
\begin{align}
	{A}_{\psi_s\bar{\psi}_s\gamma\gamma}^{(\lambda_1\lambda_2\lambda_3\lambda_4)}
	&= 
	\begin{tikzpicture}[baseline=(v1)]
	\begin{feynman}[inline = (base.a), horizontal=v1 to v1b]
		\vertex [label=\({\scriptstyle \psi_s,\,p_1}\)](a);
		\vertex [below right = 0.75 of a] (v1);
		\vertex [below left = 0.75 of v1,label=270:\({\scriptstyle \bar{\psi}_s,\,p_2}\)] (b);
		\vertex [above right = 0.75 of v1,label=\({\scriptstyle \gamma,\,p_3}\)] (c);
		\vertex [below right = 0.75 of v1,label=270:\({\scriptstyle \gamma,\,p_4}\)] (d);
		\node [left = -0.25cm of v1, blob, fill=gray] (v1b);
		\begin{pgfonlayer}{bg}
		\diagram*{
		(a) -- (v1) -- (b),
		(a) -- [photon] (v1) -- [photon] (b),
		(d) -- [photon] (v1) -- [photon] (c),
		};
		\end{pgfonlayer}
	\end{feynman}
	\end{tikzpicture},
\end{align}
and take all the momentum incoming. 
The helicity is denoted by $\lambda_i$.
The amplitude has the poles, where the intermediate spin-$s$ particles go on-shell, at
\begin{align}
	\hat{p}_{13}^2 = m^2,
	\quad
	\hat{p}_{14}^2 = m^2.
	\label{eq:pole_photon}
\end{align}
Here we denote $p_{ij} = p_i + p_j$ and use a ``hat" notation for variables deformed under the ALT shift.
We thus obtain
\begin{align}
	{A}_{\psi_s\bar{\psi}_s\gamma\gamma}^{(\lambda_1\lambda_2\lambda_3\lambda_4)}
	&= \sum_{i=3,4}\frac{1}{p_{1i}^2 - m^2}\frac{1}{z_{1i}^+ - z_{1i}^-}
	\sum_\lambda
	\left[z_{1i}^+ \hat{A}_{\psi_s \bar{\psi}_s\gamma}^{(\lambda_1 \lambda \lambda_i)}(z_{1i}^-)
	\times \hat{A}_{\psi_s \bar{\psi}_s\gamma}^{(\bar{\lambda} \lambda_2 \lambda_j)}(z_{1i}^-)
	- (z_{1i}^+ \leftrightarrow z_{1i}^-)
	\right]
	+ B_\infty,
    \label{eq:sum_pole}
\end{align}
where $j$ labels the remaining leg, $j \neq i, 1,2$ and $\bar{\lambda}$ is the little group conjugate of $\lambda$ of the intermediate factorized state. Here, $z_{1i}^{\pm}$ are the two solutions of
Eq.~\eqref{eq:pole_photon}, and we sum over all the polarizations of the intermediate particle.
We can set $B_\infty = 0$ for $s \leq 3/2$ as we argued, but for now we take $s$ arbitrary and calculate the factorized part.

%%%%%%%%%%%
\subsubsection*{General spin-$s$: same helicity case}
We begin with the same helicity photon case, $\lambda_3 = \lambda_4 = +$.
In this case, the product of the three-point functions is given by
\begin{align}
	\frac{1}{t_{13}}\sum_{\lambda} \hat{A}_{\psi_s \bar{\psi}_s\gamma}^{(\lambda_1 \lambda +)}(z_{13}^\pm)
	\times \hat{A}_{\psi_s \bar{\psi}_s\gamma}^{(\bar{\lambda} \lambda_2 +)}(z_{13}^\pm)
	&= \frac{(-1)^{2s}}{m^{2s-2}t_{13}}\hat{x}_{1I}\hat{x}_{I2} \langle \mathbf{12}\rangle^{2s} = (-)^{2s+1}\frac{\langle \mathbf{12}\rangle^{2s} [34]^2}{t_{13}\hat{t}_{14} m^{2s-2}},
\end{align}
where we denote $t_{ij} = 2p_i \cdot p_j$.
 Note that the numerator is invariant under the ALT shift. Similarly, 
\begin{align}
	\frac{1}{t_{14}}\sum_{\lambda} \hat{A}_{\psi_s \bar{\psi}_s\gamma}^{(\lambda_1 \lambda +)}(z_{14}^\pm)
	\times \hat{A}_{\psi_s \bar{\psi}_s\gamma}^{(\bar{\lambda} \lambda_2 +)}(z_{14}^\pm)
	&=  (-)^{2s+1}\frac{\langle \mathbf{12}\rangle^{2s} [34]^2}{\hat{t}_{13} t_{14} m^{2s-2}} \ .
\end{align}
Eqs.~\eqref{eq:t-channel_photon} and~\eqref{eq:u-channel_photon} are 
of the form that the factorization of one channel contains the pole of the other channel. 
One can then bring the factorized part of the amplitude as
\begin{align}
	\left. {A}_{\psi_s\bar{\psi}_s\gamma\gamma}^{(\lambda_1\lambda_2++)}\right\vert_{\mathrm{fact.}}
	&= (-)^{2s+1}\frac{\langle \mathbf{12}\rangle^{2s} [34]^2}{m^{2s-2}}\sum\mathrm{Res}_{z=z_{13}^\pm, z_{14}^\pm}\left[-
	\frac{1}{z}\frac{1}{\hat{t}_{13} \hat{t}_{14}} 
	\right].
\end{align}
The term in the square bracket has poles at $z = 0, z_{13}^\pm, z_{14}^\pm$ and vanishes sufficiently fast at $\vert z \vert \to \infty$.
It then follows from the Cauchy theorem that the above sum is equivalent to minus the residue of the pole at $z = 0$, 
and therefore we obtain
%%%%%%
\begin{align}
	{A}_{\psi_s\bar{\psi}_s\gamma\gamma}^{(\lambda_1\lambda_2++)}
	&= (-)^{2s+1} \frac{\langle \mathbf{12}\rangle^{2s} [34]^2}{ m^{2s-2}{t}_{13}{t}_{14}} + B_\infty.
\label{samehelphoton}\end{align}
%%%%%%
The factorized part is manifestly little group covariant and independent of the shift momentum $q_i$ for an arbitrary $s$,
despite that setting $B_\infty = 0$ is proven only for $s \leq 3/2$ and not for $s > 3/2$. 
 Indeed, the factorized part reproduces the result in~\cite{Aoude:2020onz}.
 
%
%%%%%%%%%%%%%%%%%
\subsubsection*{General spin-$s$: opposite helicity case}

We now turn to the opposite helicity case, $\lambda_3 = -$ and $\lambda_4 = +$.
This channel contains a spurious pole in the naive gluing method.
We now see that, with the ALT shift, the factorized part of the amplitude is free from the spurious pole.

The factorized parts in Eq.~\eqref{eq:sum_pole} are given by
%%%
\begin{align}
 \sum_{\lambda} \hat{A}_{\psi_s \bar{\psi}_s\gamma}^{(\lambda_1 \lambda -)}(z_{13}^\pm)
	\times \hat{A}_{\psi_s \bar{\psi}_s\gamma}^{(\bar{\lambda} \lambda_2 +)}(z_{13}^\pm)
	 &= \hat{\tilde{x}}_{1I} {\hat{x}}_{I2} \frac{([\textbf{1}|\hat{p}_{13}|\textbf{2}\rangle)^{2s}}{m^{4s-2}} 
    = -\frac{([4|\hat{p}_1 | 3\rangle)^2([\textbf{1}|\hat{p}_{13}|\textbf{2}\rangle)^{2s}}{\hat{t}_{14}m^{4s}} \ ,
     \label{eq:t-channel_photon}
\end{align}
%%%%%%%%%%
and
%%%%%%%%
\begin{align}
 \sum_{\lambda} \hat{A}_{\psi_s \bar{\psi}_s\gamma}^{(\lambda_1 \lambda -)}(z_{14}^\pm)
	\times \hat{A}_{\psi_s \bar{\psi}_s\gamma}^{(\bar{\lambda} \lambda_2 +)}(z_{14}^\pm) &=
	-\frac{([4|\hat{p}_1 | 3\rangle)^2(\langle \textbf{1}|\hat{p}_{14}|\textbf{2}])^{2s}}{\hat{t}_{13}m^{4s}} \ .
    \label{eq:u-channel_photon}
\end{align}
%%%%%%%%%%%%%%
Using the Schouten identity and the pole condition, we can rewrite them as
\begin{align}
 \frac{1}{t_{13}}\sum_{\lambda} \hat{A}_{\psi_s \bar{\psi}_s\gamma}^{(\lambda_1 \lambda -)}(z_{13}^\pm)
	\times \hat{A}_{\psi_s \bar{\psi}_s\gamma}^{(\bar{\lambda} \lambda_2 +)}(z_{13}^\pm)
	 &=	(-)^{2s+1}\frac{\left(\langle \mathbf{1}3\rangle [\mathbf{2}4] + [\mathbf{1}4] \langle\mathbf{2}3\rangle \right)^{2s}}{t_{13}\hat{t}_{14}[ 4\vert \hat{p}_{1}\vert 3\rangle^{2s-2}}  \ ,
	 \label{eq:sp_pole1}
     \\ 
      \frac{1}{t_{14}}\sum_{\lambda} \hat{A}_{\psi_s \bar{\psi}_s\gamma}^{(\lambda_1 \lambda +)}(z_{14}^\pm)
	\times \hat{A}_{\psi_s \bar{\psi}_s\gamma}^{(\bar{\lambda} \lambda_2 -)}(z_{14}^\pm)
	 &=(-)^{2s+1}	\frac{\left(\langle \mathbf{1}3\rangle [ \mathbf{2}4] + [\mathbf{1}4] \langle \mathbf{2}3 \rangle \right)^{2s}}{\hat{t}_{13}t_{14}[ 4\vert \hat{p}_{1}\vert 3\rangle^{2s-2}} \ .
     \label{eq:sp_pole2}
\end{align}
%%%%
We can bring the sum of the poles into the form
\begin{align}
	{A}_{\psi_{s}\bar{\psi}_{s}\gamma\gamma}^{(\lambda_1\lambda_2-+)}
	&= (-)^{2s+1}\left(\langle \mathbf{1}3\rangle [ \mathbf{2}4] + [\mathbf{1}4] \langle \mathbf{2}3 \rangle\right)^{2s}
	\times \sum \mathrm{Res}_{z=z_{13}^\pm, z_{14}^\pm}\left[-\frac{1}{z}\frac{1}{[ 4\vert \hat{p}_{1}\vert 3\rangle^{2s-2}\hat{t}_{13}\hat{t}_{14}} \right]
	+ B_\infty.
\end{align}
The term in the bracket has the poles at $z = 0, z_{13}^\pm, z_{14}^\pm$ for any $s$,
and in addition at $z = z_\mathrm{sp}$ where $[ 4\vert \hat{p}_1 \vert 3\rangle = 0$ for $s > 1$.
It vanishes faster than $1/z$ at $\vert z \vert \to \infty$, and therefore the sum over the poles $z_{13}^\pm, z_{14}^\pm$
results in
\begin{align}
	{A}_{\psi_{s}\bar{\psi}_{s}\gamma\gamma}^{(\lambda_1\lambda_2-+)}
	&= (-)^{2s+1}\left(\langle \mathbf{1}3\rangle [ \mathbf{2}4] + [\mathbf{1}4] \langle \mathbf{2}3 \rangle \right)^{2s}
	\frac{[ 4\vert p_{1}\vert 3\rangle^{2-2s}}{{t}_{13}{t}_{14}},
\end{align}
for $ 0 \leq s \leq 1$ where we set $B_\infty = 0$ proven by our (dimensional) analysis on constructibility, and
\begin{align}
	{A}_{\psi_{s}\bar{\psi}_{s}\gamma\gamma}^{(\lambda_1\lambda_2-+)}
	&= (-)^{2s+1}\frac{\left(\langle \mathbf{1}3\rangle [ \mathbf{2}4] + [\mathbf{1}4] \langle \mathbf{2}3 \rangle \right)^{2s}}{[ 4\vert p_{1}\vert 3\rangle^{2s-2}}
	\left[\frac{1}{{t}_{13}{t}_{14}} +\frac{(-z_\mathrm{sp})^{2s-2}}{(2s-3)!}\frac{d^{2s-3}}{dz^{2s-3}}
	\left(\frac{1}{z}\frac{1}{\hat{t}_{13}\hat{t}_{14}}\right)_{z\to z_\mathrm{sp}} \right]
	+ B_\infty,
\end{align}
for $s > 1$, where we used
\begin{align}
	\frac{1}{[ 4\vert \hat{p}_{1}\vert 3\rangle}
	= -\frac{z_\mathrm{sp}}{z-z_\mathrm{sp}}\frac{1}{[ 4\vert p_{1}\vert 3\rangle}.
\end{align}
In the following, we focus on $s = 3/2$, for which we can set $B_\infty = 0$, and simplify the above expression. 

%%%%%%%%%%%%%%%%%%%%%%%%%%%%%%
\subsubsection{Spin-3/2 Amplitude}
%%%%%%%%%%%%%%%%%%%%%%%%%%%%%%
We now set $s = 3/2$ and simplify the expression we obtained above, which reduces to
\begin{align}
	{A}_{\psi_{3/2}\bar{\psi}_{3/2}\gamma\gamma}^{(\lambda_1\lambda_2-+)}
	&= \frac{\left(\langle \mathbf{1}3\rangle [ \mathbf{2}4] + [\mathbf{1}4] \langle \mathbf{2}3 \rangle \right)^{3}}{[ 4\vert p_{1}\vert 3\rangle}
	\left[
	\frac{1}{t_{13}t_{14}}
	- \left.\frac{1}{\hat{t}_{13}\hat{t}_{14}}\right\vert_{z=z_\mathrm{sp}}
	\right],
\end{align}
where we set $B_\infty = 0$ as this is justified by our dimensional analysis.
This form makes it explicit that the second term cancels the spurious pole in the first term.\footnote{Ref.~\cite{Chiodaroli:2021eug} proposed subtraction of spurious poles to derive a spurious pole-free amplitude but did not specify a momentum shift scheme. }
We emphasize that this is an automatic outcome of our calculation with the explicit momentum shift,
and this should be distinguished from merely subtracting the spurious pole by hand, without a specific momentum shift, as the latter procedure has an undetermined ambiguity of additional boundary term contributions while the former does not.
The second term can be simplified by using
\begin{align}
	\langle \mathbf{1}3\rangle [ \mathbf{2}4] + [ \mathbf{1}4]\langle \mathbf{2}3 \rangle
	&= -\frac{\hat{p}_{13}^2 - m^2}{m^2}[\mathbf{1}4]\langle \mathbf{2}3 \rangle
	-\frac{[ 4\vert \hat{p}_{13} \vert 3\rangle \langle \mathbf{1}\vert \hat{p}_{13}\vert \mathbf{2}]}{m^2}
	\nonumber \\
	&= -\frac{\hat{p}_{14}^2 - m^2}{m^2}\langle \mathbf{1}3 \rangle [ \mathbf{2}4 ] - \frac{[ 4\vert \hat{p}_{14}\vert 3 \rangle \langle \mathbf{2}\vert \hat{p}_{14}\vert \mathbf{1}]}{m^2},
	\label{eq:sp_pole_identity}
\end{align}
which hold for an arbitrary $z$. In particular, the second terms vanish at $z = z_\mathrm{sp}$, and we obtain
\begin{align}
	{A}_{\psi_{3/2}\bar{\psi}_{3/2}\gamma\gamma}^{(\lambda_1\lambda_2-+)}
	&= \frac{\langle \mathbf{1}3\rangle [ \mathbf{2}4] + [\mathbf{1}4] \langle \mathbf{2}3 \rangle}{[4\vert p_1 \vert 3 \rangle }
	\left[\frac{\left(\langle \mathbf{1}3\rangle [ \mathbf{2}4] + [\mathbf{1}4] \langle \mathbf{2}3 \rangle \right)^2}
	{t_{13}t_{14}}
	- \frac{[ \mathbf{1}4] \langle \mathbf{2}3 \rangle \langle \mathbf{1}3\rangle [ \mathbf{2}4]}{m^4}
	\right],
    \label{eq:four-point-compton-three-half}
\end{align}
which is independent of $q_i$ as expected.
The first term corresponds to the amplitude one may obtain by a naive gluing, which has a spurious pole
due to the prefactor $1/[ 4\vert p_1 \vert 3\rangle$, canceled in our result by the second term.
This result is consistent with Ref.~\cite{Chiodaroli:2021eug}, where a spurious pole subtraction was conjectured and utilized.

Although concise, the above expression contains the prefactor $1/[4\vert p_1 \vert 3\rangle$.
Since our result is free from the spurious pole, we can explicitly eliminate this prefactor from the amplitude.
For this purpose, we may go back to Eqs.~\eqref{eq:t-channel_photon} and~\eqref{eq:u-channel_photon}
and bring them to the form
\begin{equation}
\begin{aligned}
 \sum_{\lambda} \hat{A}_{\psi_s \bar{\psi}_s\gamma}^{(\lambda_1 \lambda +)}(z_{13}^\pm)
	\times \hat{A}_{\psi_s \bar{\psi}_s\gamma}^{(\bar{\lambda} \lambda_2 -)}(z_{13}^\pm)
	 &=\frac{\hat{f}_{tu}}{\hat{t}_{14}} + \hat{f}_t \ , \\ 
      \sum_{\lambda} \hat{A}_{\psi_s \bar{\psi}_s\gamma}^{(\lambda_1 \lambda +)}(z_{14}^\pm)
	\times \hat{A}_{\psi_s \bar{\psi}_s\gamma}^{(\bar{\lambda} \lambda_2 -)}(z_{14}^\pm)
	 &=\frac{\hat{f}_{tu}}{\hat{t}_{13}} + \hat{f}_u \ .
\end{aligned}
    \label{eq:cc}
\end{equation}
We introduce the following shorthand notations to represent recurrent spinor structures:
%%%%
\begin{align}
   \mathcal{U} = [\textbf{2}4] \langle\textbf{1}3 \rangle, ~~ \mathcal{V} = [\textbf{1}4] \langle\textbf{2}3 \rangle,
   ~~ \mathcal{X} = [\textbf{1}\lvert p_{13}\lvert \textbf{2}\rangle,~~ \mathcal{Y} =[\textbf{2}\lvert p_{14}\lvert \textbf{1}\rangle,
  ~~\mathcal{T} = \mathcal{U} + \mathcal{V},~~ \mathcal{S}= \mathcal{X} + \mathcal{Y}.
\end{align}
%%%%
The ALT shift deforms ${\mathcal{X}}$ and ${\mathcal{Y}}$ 
due to the momentum insertion as $\hat{\mathcal{X}}$ and $\hat{\mathcal{Y}}$,
while it does not modify $\mathcal{U}$ and $\mathcal{V}$.
By setting $s = 3/2$ in Eq.~\eqref{eq:t-channel_photon}, we obtain
%%%%%%%
\begin{align}
 \sum_{\lambda} \hat{A}_{\psi_{3/2} \bar{\psi}_{3/2}\gamma}^{(\lambda_1 \lambda -)}(z_{13}^\pm)
	\times \hat{A}_{\psi_{3/2} \bar{\psi}_{3/2}\gamma}^{(\bar{\lambda} \lambda_2 +)}(z_{13}^\pm)
    =- \frac{([4|\hat{p}_1 | 3\rangle)^2([\textbf{1}|\hat{p}_{13}|\textbf{2}\rangle)^{3}}{\hat{t}_{14}m^{6}} 
    &= -  \frac{([4|\hat{p}_1 | 3\rangle)^2(\hat{\mathcal{S}} - [\textbf{1}\lvert \frac{\hat{p}_3 \hat{p}_4}{m}\lvert \textbf{2}])^{3}}{2^3 \hat{t}_{14}m^{6}} \ .
     \label{eq:t-channel_photon1_threehalf}
\end{align}
%%%%%%%%%%
We apply the following identity, which holds at the pole $\hat{p}^2_{13}= m^2$:
%%%%%%%
\begin{align}
  &  [4|\hat{p}_1 | 3\rangle \hat{\mathcal{S}} = -(\hat{t}_{14} \mathcal{U} + 2m^2 \mathcal{T}),
  \quad
    [4|\hat{p}_1 | 3\rangle [\textbf{1}\lvert \frac{\hat{p}_3 \hat{p}_4}{m}\lvert \textbf{2}] = -\hat{t}_{14} \mathcal{U} \ . \label{eq:id} 
\end{align}
%%%%%%%
We then find that
%%%%%%%
\begin{align}
 \sum_{\lambda} \hat{A}_{\psi_{3/2} \bar{\psi}_{3/2}\gamma}^{(\lambda_1 \lambda -)}(z_{13}^\pm)
	\times \hat{A}_{\psi_{3/2} \bar{\psi}_{3/2}\gamma}^{(\bar{\lambda} \lambda_2 +)}(z_{13}^\pm)
    &= - \frac{\hat{\mathcal{S}}(4m^4\mathcal{T}^2- 2m^2 \mathcal{T} \mathcal{U}\hat{t}_{14})+ 2\hat{\mathcal{X}} \mathcal{U}^2\hat{t}_{14}^2}{2^3 \hat{t}_{14}m^{6}} \ .
     \label{eq:t-channel_photon1_threehalf1}
\end{align}
%%%%%%%
The final step is to reduce one power of $\hat{t}_{14}$ from the last term.
To do so, we use
\begin{align}
    \mathcal{U} \hat{t}_{14} = - [\textbf{2}\lvert \hat{p}_4 \lvert \textbf{1} \rangle \langle 3 \lvert \hat{p}_1 \lvert 4 ] - m[\textbf{2}\lvert \hat{p}_4 \lvert 3 \rangle [\textbf{1}4] , \label{eq:id2} 
\end{align}
which follows from the Schouten identity applied to the left-hand side.
After performing spinor manipulations, we obtain
% 
%%%%%%%
\begin{align}
 \sum_{\lambda} \hat{A}_{\psi_{3/2} \bar{\psi}_{3/2}\gamma}^{(\lambda_1 \lambda -)}(z_{13}^\pm)
	\times \hat{A}_{\psi_{3/2} \bar{\psi}_{3/2}\gamma}^{(\bar{\lambda} \lambda_2 +)}(z_{13}^\pm)
    &= - \frac{\hat{\mathcal{S}} \mathcal{T}^2}{2 \hat{t}_{14}m^2} + \frac{\mathcal{U}\hat{\mathcal{X}} \mathcal{T}}{2m^4} \ .
     \label{eq:t-channel_photon1_threehalf2}
\end{align}
%%%%%%%
In a similar way, we obtain
%%%%%%%
\begin{align}
 \sum_{\lambda} \hat{A}_{\psi_{3/2} \bar{\psi}_{3/2}\gamma}^{(\lambda_1 \lambda +)}(z_{14}^\pm)
	\times \hat{A}_{\psi_{3/2} \bar{\psi}_{3/2}\gamma}^{(\bar{\lambda} \lambda_2 -)}(z_{14}^\pm)
    &= - \frac{\hat{\mathcal{S}} \mathcal{T}^2}{2 \hat{t}_{13}m^2} + \frac{\mathcal{V}\hat{\mathcal{Y}} \mathcal{T}}{2m^4} \ .
     \label{eq:t-channel_photon1_threehalf3}
\end{align} 
%%%%%%%
They are now of the form~\eqref{eq:cc}. 
The second term contains only up to terms linear in $z$.
The terms linear in $z$ cancel after the summation of $z_{1i}^\pm$, and therefore we can remove the hats from 
$\hat{\mathcal{X}}$ and $\hat{\mathcal{Y}}$.
The first terms can be combined to be written as
\begin{align}
	\frac{\mathcal{T}^2}{2m^2}
	\times \sum \mathrm{Res}_{z=z_{13}^\pm, z_{14}^\pm} \left[\frac{1}{z}\frac{\hat{\mathcal{S}}}{\hat{t}_{13}\hat{t}_{14}}\right].
\end{align}
The term in the square bracket has the poles at $z = 0, z_{13}^\pm, z_{14}^\pm$, and vanishes sufficiently fast at $\vert z \vert \to \infty$,
and therefore this summation is equivalent to minus the residue of the pole at $z = 0$.
Thus, we obtain the opposite-helicity Compton amplitude as 
% %%%%%%
\begin{tcolorbox}[colback=gray!5, colframe=black, boxrule=0.2mm]
\begin{align}
    {A}_{\psi_{3/2}\bar{\psi}_{3/2}\gamma\gamma}^{(\lambda_1\lambda_2-+)}
    = & - \frac{\mathcal{S} \mathcal{T}^2}{2m^2 t_{13} t_{14}} 
    + \frac{\mathcal{U} \mathcal{X} \mathcal{T}}{2m^4 t_{13}} 
    + \frac{\mathcal{V} \mathcal{Y} \mathcal{T}}{2m^4 t_{14}} \nonumber \\
     =& - \frac{\big( [\mathbf{1} | p_{13} | \mathbf{2} \rangle 
    + \langle \mathbf{1} | p_{14} | \mathbf{2} ] \big) 
    \big( [\mathbf{1}4] \langle \mathbf{2}3 \rangle 
    + [\mathbf{2}4] \langle \mathbf{1}3 \rangle \big)^2}{2m^2 t_{13} t_{14}} \nonumber \\
    & + \frac{\big([\mathbf{1}4] \langle \mathbf{2}3 \rangle 
    + [\mathbf{2}4] \langle \mathbf{1}3 \rangle \big)}{2m^4}
    \left[\frac{[\mathbf{2}4] \langle \mathbf{1}3 \rangle 
    [\mathbf{1} | p_{13} | \mathbf{2} \rangle 
    }{t_{13}}
    + \frac{[\mathbf{1}4] \langle \mathbf{2}3 \rangle 
    \langle \mathbf{1} | p_{14} | \mathbf{2} ]}{t_{14}}
    \right] \ .
    \label{eq:four-Compton-three-half}
\end{align}
\end{tcolorbox}
\noindent
This result is explicitly free from spurious poles. Note again that for on-shell constructible amplitudes, the factorized amplitude is free from shifted momentum $q_i$, as it should be. One can check that Eqs.~\eqref{eq:four-point-compton-three-half} and~\eqref{eq:four-Compton-three-half} are equivalent, as they should. Finally, we note that the amplitude exhibits $1/m^4$ divergence. This divergence arises from the external polarization vectors and the coupling constant, and not from the kinematic factor $F$, consistent with the assumption in our large-$z$ analysis. Note that in the minimal amplitude, we fix the scale of the dimension-five operator $\Lambda$ as $\Lambda = m$,
	contributing to the mass divergence.

  \begin{table}[h]
\centering
\renewcommand{\arraystretch}{1.4} 
\large 

\begin{minipage}{0.45\textwidth}
\centering
\begin{tabular}{|c|c|c|c|c|}
\hline
\multicolumn{5}{|c|}{\((h_3, h_4) = (-, +)\)} \\
\hline
\diagbox{$\lambda_2$}{$\lambda_1$} & \(+\frac{3}{2}\) & \(+\frac{1}{2}\) & \(-\frac{1}{2}\) & \(-\frac{3}{2}\) \\
\hline
\(+\frac{3}{2}\)  & \(E^{-3}\) & \(E^{2}\) & \(E^{3}\) & \(E^{2}\) \\
\hline
\(+\frac{1}{2}\)  & \(E^{-2}\) & \(E^{3}\) & \(E^{4}\) & \(E^{3}\) \\
\hline
\(-\frac{1}{2}\) & \(E^{-1}\) & \(E^{4}\) & \(E^{3}\) & \(E^{2}\) \\
\hline
\(-\frac{3}{2}\) & \(E^{0}\)  & \(E^{-1}\) & \(E^{-2}\) & \(E^{-3}\) \\
\hline
\end{tabular}
\end{minipage}
\hspace{1cm} 
\begin{minipage}{0.45\textwidth}
\centering
\begin{tabular}{|c|c|c|c|c|}
\hline
\multicolumn{5}{|c|}{\((h_3, h_4) = (+, +)\)} \\
\hline
\diagbox{$\lambda_2$}{$\lambda_1$} & \(+\frac{3}{2}\) & \(+\frac{1}{2}\) & \(-\frac{1}{2}\) & \(-\frac{3}{2}\) \\
\hline
\(+\frac{3}{2}\)  & \(E^{-5}\) & \(E^{-4}\) & \(E^{-3}\) & \(E^{-2}\) \\
\hline
\(+\frac{1}{2}\)  & \(E^{-4}\) & \(E^{-3}\) & \(E^{-2}\) & \(E^{-1}\) \\
\hline
\(-\frac{1}{2}\) & \(E^{-3}\) & \(E^{-2}\) & \(E^{-1}\) & \(E^{0}\) \\
\hline
\(-\frac{3}{2}\) & \(E^{-2}\)  & \(E^{-1}\) & \(E^{0}\) & \(E^{1}\) \\
\hline
\end{tabular}
\end{minipage}

  \caption{Energy dependence of different helicity configurations for spin-$3/2$ electromagnetic Compton amplitude in the CM frame. The CM frame is defined as: $p_1 = (\sqrt{\textbf{p}^2 +m^2},\textbf{p},0,0 )$, $p_2 = (-\sqrt{\textbf{p}^2 +m^2},-\textbf{p}\cos \theta,-\textbf{p}\sin \theta,0  )$, $p_3 = (\textbf{p},-\textbf{p},0,0 )$, $p_4 = (-\textbf{p},\textbf{p}\cos \theta,\textbf{p} \sin \theta,0) $, where the CM energy $E$ is given by $\textbf{p}=\frac{E^2-m^2}{2E}$.  
  } 
\label{tab:helicity_energy_spin_three_half}
\end{table}
Table~\ref{tab:helicity_energy_spin_three_half} summarizes the high-energy behavior of various helicity amplitudes in the 
center-of-mass (CM) frame, contrasting the same and opposite photon helicity configurations.\footnote{
%%%
Note that the choice of CM frame guarantees well-defined helicities for all particles.
} As we can see, the energy growth can be as fast as  $E^4$, which breaks unitarity at a relatively low cutoff scale. 
Here, we have considered a theory containing only the spin-3/2 particle and photon, 
and the dimension of the couplings is limited to be less than or equal to 5. To unitarize the amplitudes at high energy, one may need to extend the discussion and include other degrees of freedom or higher-dimensional couplings. In this case, the on-shell amplitude computation requires a re-evaluation of the large-$z$ behavior. We leave the on-shell exploration of good UV theories for future work. 

Now we return to the pattern of high-energy behavior in the helicity amplitudes. For the same photon helicity case ($h_3, h_4=+, +$), corresponding to the right panel of Table~\ref{tab:helicity_energy_spin_three_half}, the numerator of the amplitude is proportional to $\langle \textbf{1}\textbf{2}\rangle^3 [34]^2$. For the helicity configuration $(-\frac{3}{2},-\frac{3}{2})$, this expression scales as $E^5$. This piece corresponds to one of the Maximally Helicity Violating (MHV) amplitudes in high-energy. Including the denominator contribution from $t_{13} t_{14}$, the full amplitude behaves as $E^1$. Increasing the helicity of any of the massive legs replaces a $\lvert i \rangle$ spinor by a $\lvert \eta_i \rangle$ spinor, which introduces a suppression factor of $m/E$, and explains the rest of the patterns in the table. Now consider the opposite photon helicity case ($h_3, h_4=-, +$) corresponding to the left panel of Table~\ref{tab:helicity_energy_spin_three_half}. Here, the leading high-energy contribution comes from the helicity configuration $(\pm \frac{1}{2},\mp \frac{1}{2})$, and this behavior is governed by the second term in the bracket of Eq.~\eqref{eq:four-point-compton-three-half}. For these configurations, the terms contain a component without any $\eta$ spinors and scale as $E^4$. Moving up or to the right of these entries in the table again introduces an additional $\eta$ spinor, reducing the energy scaling by a factor of $m/E$ for each replacement. Note that the high-energy behavior in the first column and last row of the table differs from the rest. This can be explained by our choice of CM frame, where $\langle \eta_1 3 \rangle = [\eta_2 4] = 0$. As a result, the second term in the bracket of Eq.~\eqref{eq:four-point-compton-three-half} vanishes when the particle helicity $\lambda_1 = \frac{3}{2}$ or $\lambda_2 = -\frac{3}{2}$, and so the high-energy behavior of these entries in the table is governed solely by the first term. Again, the $E^0$ contribution for another MHV configuration ($+\frac{3}{2},-\frac{3}{2}$) comes from an $\eta$ independent component in the amplitude and the rest of the entries are related by $m/E$ suppression from helicity flip.

%

%%%%%%%%%%%%%%%%%%%
\subsubsection*{Non-minimal amplitude}
%%%%%%%%%%%%%%%%%%%

So far, we have constructed the four-point amplitude with the minimal three-point amplitudes as the input and have proven the on-shell constructibility with them.
However, on-shell constructibility is not guaranteed for the non-minimal three-point amplitudes.
Therefore, it is interesting to calculate explicitly the factorized part in the non-minimal case and explore features of the resulting amplitudes and their relation to constructibility.
We keep up to dimension-five operators, where the amplitudes are parametrized by two numbers $l_1$ and $l_2$ as
%%%%%%%%%%%%%%%%%%
\begin{align}
   &{A}_3(\psi_1^{3/2},\bar{\psi}_2^{3/2},A^{-}_3) = \frac{\tilde{x}_{12}}{m^2}(l_1 [\textbf{1}\textbf{2}]^3 + 2(2l_2 +1)[\textbf{1}\textbf{2}]^2 \langle \textbf{1}\textbf{2}\rangle-(l_1+4l_2)[\textbf{1}\textbf{2}] \langle \textbf{1}\textbf{2}\rangle^2 ) ,  \\
   &{A}_3(\psi_1^{3/2},\bar{\psi}_2^{3/2},A^{+}_3) = \frac{x_{12}}{m^2}(l_1 \langle \textbf{1}\textbf{2}\rangle^3 + 2(2l_2 +1)[\textbf{1}\textbf{2}] \langle \textbf{1}\textbf{2}\rangle^2-(l_1+4l_2)[\textbf{1}\textbf{2}]^2 \langle \textbf{1}\textbf{2}\rangle ).
\label{eq:non-minimal-three-pt}
\end{align}
%%%%%%%%%%%%%%%%%%%%%%%%
We have absorbed an overall normalization into $m$, and
$l_1$ and $l_2$ correspond the coefficients of the dimension five operators (see more details in Appendix~\ref{app:threepoint32}). 
The minimal amplitude corresponds to $l_1 = 2$ and $l_2 = -1/2$, in which case only the first term survives.

We first consider $l_1 + 4l_2 = 0$ so that the last term vanishes. The three-point amplitudes are then given by
%%%%%%%%%%%%%%%%%%
\begin{align}
   & {A}_3(\psi_1^{3/2},\bar{\psi}_2^{3/2},A^{-}_3) = \frac{\tilde{x}_{12}}{m^2}(l_1 [\textbf{1}\textbf{2}]^3 + 2(2l_2 +1)[\textbf{1}\textbf{2}]^2 \langle \textbf{1}\textbf{2}\rangle)  , \\
   & {A}_3(\psi_1^{3/2},\bar{\psi}_2^{3/2},A^{+}_3) = \frac{x_{12}}{m^2}(l_1 \langle \textbf{1}\textbf{2}\rangle^3 + 2(2l_2 +1)[\textbf{1}\textbf{2}] \langle \textbf{1}\textbf{2}\rangle^2 )  .
\end{align}
%%%%%%%%%%%%%%%%%%%%%%%%
The factorized part involves four distinct spinor contractions, i.e.,
\begin{align}
    (a)\  [\textbf{1}\textbf{I}]^3 \otimes \langle -\textbf{I}\textbf{2}\rangle^3, \quad (b)\ [\textbf{1}\textbf{I}]^2 \langle \textbf{1} \textbf{I}\rangle \otimes [-\textbf{I}\textbf{2}] \langle -\textbf{I}\textbf{2}\rangle^2, \quad (c) \ [\textbf{1}\textbf{I}]^3 \otimes [-\textbf{I}\textbf{2}] \langle -\textbf{I}\textbf{2}\rangle^2, \quad (d) \ [\textbf{1}\textbf{I}]^2 \langle \textbf{1} \textbf{I}\rangle \otimes \langle -\textbf{I}\textbf{2}\rangle^3,
\end{align}
%%%%%%%
for the $t$-channel ($\hat{t}_{13}=0$).
%%%%%%%
Here, $\mathbf{I}$ stands for the intermediate spin-3/2 particle, whose helicity is summed over (represented by the operator $\otimes$). Structure $(a)$ corresponds to the minimal amplitude we computed in Eq.~\eqref{eq:four-Compton-three-half}. Only structure $(b)$ remains in the case $l_1 = l_2 = 0$ and it involves only the dimension-four operator. 
In this case, we have
%%%%%%%%
\begin{align}
 \sum_{\lambda} \hat{A}_{\psi_{3/2} \bar{\psi}_{3/2}\gamma}^{(\lambda_1 \lambda -)}
	\times \hat{A}_{\psi_{3/2} \bar{\psi}_{3/2}\gamma}^{(\bar{\lambda} \lambda_2 +)}\bigg|_{z^{\pm}_{13}}^{l_1=l_2=0} &= - \frac{[4\lvert \hat{p}_1 \lvert 3\rangle^2  }{\hat{t}_{14}m^6} \bigg(\frac{1}{3}[\textbf{1}\lvert \hat{p}_{13}\lvert \textbf{2}\rangle^2[\textbf{2}\lvert \hat{p}_{13}\vert \textbf{1}\rangle + \frac{2}{3}m^2 [\textbf{1}\textbf{2}]\langle \textbf{1} \textbf{2} \rangle [\textbf{1}\lvert \hat{p}_{13}\lvert \textbf{2}\rangle \bigg) \nonumber \\
    &=  \frac{[4\lvert \hat{p}_1 \lvert 3\rangle  }{\hat{t}_{14}m^2} {\mathcal{T}} \bigg( [\textbf{1} \textbf{2}]\langle\textbf{1} \textbf{2} \rangle- \frac{1}{3} \frac{[\textbf{1}\lvert \hat{p}_{3}\lvert \textbf{2} \rangle \langle \textbf{1}\lvert \hat{p}_{4}\lvert \textbf{2} ]}{m^2}\bigg) - \frac{1}{3} \frac{{\mathcal{T}}{\mathcal{U}} \langle \textbf{1}\textbf{2}\rangle}{m^3} \ , 
    \\
    \sum_{\lambda} \hat{A}_{\psi_{3/2} \bar{\psi}_{3/2}\gamma}^{(\lambda_1 \lambda +)}
	\times \hat{A}_{\psi_{3/2} \bar{\psi}_{3/2}\gamma}^{(\bar{\lambda} \lambda_2 -)}\bigg|_{z_{14}^{\pm}}^{l_1=l_2=0}&=  - \frac{[4\lvert \hat{p}_1 \lvert 3\rangle^2  }{\hat{t}_{13}m^6}  \bigg( \frac{1}{3} [\textbf{1}\lvert \hat{p}_{14}\lvert \textbf{2}\rangle [\textbf{2}\lvert \hat{p}_{14}\vert \textbf{1}\rangle^2 + \frac{2}{3} [\textbf{1}\textbf{2}] \langle \textbf{1} \textbf{2} \rangle  \langle \textbf{1}\lvert \hat{p}_{14}\lvert \textbf{2}]  \bigg) \nonumber \\
    &=  \frac{[4\lvert \hat{p}_1 \lvert 3\rangle  }{\hat{t}_{13}m^2} {\mathcal{T}} \bigg( [\textbf{1} \textbf{2}]\langle\textbf{1} \textbf{2} \rangle- \frac{1}{3} \frac{[\textbf{1}\lvert \hat{p}_{3}\lvert \textbf{2} \rangle \langle \textbf{1}\lvert \hat{p}_{4}\lvert \textbf{2} ]}{m^2}\bigg) - \frac{1}{3}  \frac{{\mathcal{T}}{\mathcal{V}} [\textbf{1}\textbf{2}]}{m^3} \ .
\end{align}
%%%%%%%%
%%%%%%%%
This is of the form~Eq.~\eqref{eq:cc} and does not generate a contact term. 
The sum over all the physical poles gives us
%%%%%%%%%%%
\begin{align}
    A_4^{(b)} =   \frac{[4\lvert {p}_1 \lvert 3\rangle  }{{t}_{14}{t}_{13}m^2} {\mathcal{T}} \bigg( [\textbf{1} \textbf{2}]\langle\textbf{1} \textbf{2} \rangle-\frac{1}{3} \frac{[\textbf{1}\lvert {p}_{3}\lvert \textbf{2} \rangle \langle \textbf{1}\lvert {p}_{4}\lvert \textbf{2}]}{m^2}\bigg) - \frac{1}{3} \frac{{\mathcal{T}}{\mathcal{U}} \langle \textbf{1}\textbf{2}\rangle}{m^3{t}_{13}} -  \frac{1}{3} \frac{{\mathcal{T}}{\mathcal{V}} [\textbf{1}\textbf{2}]}{m^3{t}_{14}} \ .
\end{align}
%%%%%%%%%%%
The contribution from $(c)$ and $(d)$ takes the form
%%%%%%%
\begin{align}
 \sum_{\lambda} \hat{A}_{\psi_{3/2} \bar{\psi}_{3/2}\gamma}^{(\lambda_1 \lambda -)}
	\times \hat{A}_{\psi_{3/2} \bar{\psi}_{3/2}\gamma}^{(\bar{\lambda} \lambda_2 +)}\bigg|_{z^{\pm}_{13}}^{(c+d)} &=  \frac{[4\lvert \hat{p}_1 \lvert 3\rangle^2  }{\hat{t}_{14}m^5} [\textbf{1}\lvert \hat{p}_{13}\lvert \textbf{2}\rangle^2([\textbf{1}\textbf{2}]+\langle\textbf{1}\textbf{2} \rangle) 
	%\nonumber \\
    =  \frac{{\mathcal{T}}^2 }{\hat{t}_{14}m} ([\textbf{1}\textbf{2}]+\langle\textbf{1}\textbf{2} \rangle) \ , \\
      \sum_{\lambda} \hat{A}_{\psi_{3/2} \bar{\psi}_{3/2}\gamma}^{(\lambda_1 \lambda +)}
	\times \hat{A}_{\psi_{3/2} \bar{\psi}_{3/2}\gamma}^{(\bar{\lambda} \lambda_2 -)}\bigg|_{z_{14}^{\pm}}^{(c+d)} 
	&=  \frac{[4\lvert \hat{p}_1 \lvert 3\rangle^2  }{\hat{t}_{13}m^5} [\textbf{2}\lvert \hat{p}_{14}\lvert \textbf{1}\rangle^2([\textbf{1}\textbf{2}]+\langle\textbf{1}\textbf{2} \rangle) 
	%\nonumber \\
    =   \frac{{\mathcal{T}}^2 }{\hat{t}_{13}m} ([\textbf{1}\textbf{2}]+\langle\textbf{1}\textbf{2} \rangle) \ .
\end{align}
%%%%%%%%%%%
After the summation over poles, we obtain
\begin{align}
    A_4^{(c+d)} =   \frac{{\mathcal{T}}^2 }{{t}_{14}{t}_{13}m} ([\textbf{1}\textbf{2}]+\langle\textbf{1}\textbf{2} \rangle)  \ .
\end{align}
%%%%%%%
Therefore, the factorized part of the amplitude for $l_1 + 4l_2 =0$ is given by 
%%%%%%%%%%%%
\begin{align}
    \left.A_4^{(l_1 +4l_2 = 0)}\right\vert_{\mathrm{fact.}} = l_1^2 A_4^{(\mathrm{min})} + 4(2l_2+1)^2 A_4^{(b)} + 2 l_1(2l_2+1) A_4^{(c+d)},
\end{align}
%%%%%%%%%%%
where $A_4^{(\mathrm{min})}$ refers to amplitude in Eq.~\eqref{eq:four-Compton-three-half}. Now, we turn to the general case $l_1+4l_2 \neq 0$. 
For this, we need additional spinor contractions, i.e.,
%%%%%%%%%%%%%%
\begin{align}
    (e)\  [\textbf{1}\textbf{I}] \langle \textbf{1}\textbf{I} \rangle^2 \otimes [ -\textbf{I}\textbf{2}]^2 \langle -\textbf{I}\textbf{2} \rangle, \quad (f)\  [\textbf{1}\textbf{I}] \langle \textbf{1}\textbf{I} \rangle^2  &\otimes [-\textbf{I}\textbf{2}] \langle -\textbf{I}\textbf{2}\rangle^2, \quad (g)[\textbf{1}\textbf{I}]^2 \langle \textbf{1} \textbf{I}\rangle \otimes [-\textbf{I}\textbf{2}]^2 \langle -\textbf{I}\textbf{2}\rangle \ , \quad  \nonumber \\  (h) \  [\textbf{1}\textbf{I}]^3 \otimes [-\textbf{I}\textbf{2}]^2 \langle -\textbf{I}\textbf{2}\rangle ,& \quad   (i) \ [\textbf{1}\textbf{I}] \langle \textbf{1}\textbf{I} \rangle^2 \otimes \langle -\textbf{I}\textbf{2}\rangle^3 \ ,
\end{align}
%%%%%%%%%%%%%
 for the $t$-channel. After summing over the poles, we find that the factorized amplitude is again independent of the shifted momentum. For structure $(e)$, we get
 %%%%%
\begin{align}
    A_4^{(e)} =&   \frac{1}{3}\frac{[4\lvert {p}_1 \lvert 3\rangle \mathcal{T} }{{t}_{13}t_{14}m^2} \bigg( \frac{{\mathcal{X}} {\mathcal{Y}}}{m^2} +[\textbf{1}\textbf{2}]^2+ \langle  \textbf{1}\textbf{2} \rangle^2  \bigg) - \frac{2}{3} \frac{ ([\textbf{1}\textbf{2}]+\langle  \textbf{1}\textbf{2} \rangle)\left(\mathcal{T}^2- \frac{[4\lvert {p}_1 \lvert 3\rangle^2\langle \textbf{1}\textbf{2}\rangle [\textbf{1}\textbf{2}]}{m^2} \right) }{{t}_{13} t_{14}m}   \nonumber \\
     & - \frac{1}{3}\frac{ \mathcal{U} \mathcal{T}}{m^3t_{13}} \bigg( [\textbf{1}\textbf{2}]+\langle  \textbf{1}\textbf{2} \rangle - \frac{1}{m}\langle \textbf{1} \lvert {p}_{13}\lvert \textbf{2} ] \bigg)-\frac{2}{3}\frac{\mathcal{U}[4\lvert {p}_1 \lvert 3\rangle \langle \textbf{1}\textbf{2}\rangle [\textbf{1}\textbf{2}]}{m^4t_{13}}   \nonumber \\
    & - \frac{1}{3}\frac{ \mathcal{V} \mathcal{T}}{m^3t_{14}} \bigg( [\textbf{1}\textbf{2}]+\langle  \textbf{1}\textbf{2} \rangle - \frac{1}{m}\langle \textbf{2} \lvert {p}_{14}\lvert \textbf{1} ] \bigg)-\frac{2}{3}\frac{\mathcal{V}[4\lvert {p}_1 \lvert 3\rangle \langle \textbf{1}\textbf{2}\rangle [\textbf{1}\textbf{2}]}{m^4t_{14}}  \ . 
\end{align}
%%%%%%%%%%%%%%
For $(h)$ and $(i)$, we find
%%%%%%%%%%%%%%%%
\begin{align}
     A_4^{(h+i)} &=  \frac{[4\lvert {p}_1 \lvert 3\rangle  }{{t}_{13}t_{14}m^2} \mathcal{T}  \left( [\textbf{1}\textbf{2}]^2 + \langle \textbf{1}\textbf{2}\rangle^2   \right)   .
\end{align}
%%%%%%%%%%%%%%%%%%
%%%%%%%%%%%%%%%%%%%
Finally, we find for $(f)$ and $(g)$
%%%%%%%%%%%%%%%%%%%
\begin{align}
    A_4^{(f+g)} &= - \frac{2}{3} \frac{1}{{t}_{13} t_{14} m^2} \mathcal{T} 
    \left( \langle \mathbf{1} \mathbf{2} \rangle + [\mathbf{1} \mathbf{2}]  \right) 
    \left( m \mathcal{T} - [4| p_1 | 3\rangle \left(\langle \mathbf{1} \mathbf{2} \rangle 
    + [\mathbf{1} \mathbf{2}] \right)  \right)  \nonumber \\
    & + \frac{1}{3} \frac{[4| p_1 | 3\rangle^2}{{t}_{13} t_{14} m^3} 
    [\mathbf{1} \mathbf{2}] \langle \mathbf{1} \mathbf{2} \rangle 
    \left( \langle \mathbf{1} \mathbf{2} \rangle + [\mathbf{1} \mathbf{2}]  \right) 
    % \nonumber \\
    %&\quad 
    - \frac{2}{3} \frac{1}{t_{13} m^3} 
    \left( \langle \mathbf{1} \mathbf{2} \rangle + [\mathbf{1} \mathbf{2}]  \right) 
    \mathcal{U} \mathcal{T}  
    - \frac{2}{3} \frac{1}{t_{14} m^3} 
    \left( \langle \mathbf{1} \mathbf{2} \rangle + [\mathbf{1} \mathbf{2}]  \right) 
    \mathcal{V} \mathcal{T} \, .
\end{align}
Collecting all the results, we obtain the factorized part as
%%%%%%%%%%%%%%%%%%
\begin{align}
    \left.A_4^{(l_1 +4l_2 \neq 0)}\right\vert_{\mathrm{fact.}} 
    =  \left.A_4^{(l_1 +4l_2 = 0)}\right\vert_{\mathrm{fact.}}- (l_1+4l_2) \left(-(l_1+4l_2) A_4^{(e)}+ 2(2l_2+1)A_4^{(f+g)}+ l_1 A_4^{(h+i)}\right) . 
    \label{eq:non-minimal-four-point}
\end{align}
%%%%%%%%%%%%%%%%%%
The factorized part of the amplitude is $q_i$-independent and is little-group covariant. 
Furthermore, the amplitude has at most $1/m^4$ divergence, similar to the minimal case. 

As we have argued in Sec.~\ref{sec:constructibility} and  App.~\ref{app:threepoint32}, the non-minimal amplitudes 
with $l_1\neq 2 $ and/or $ l_2\neq-1/2$ do not satisfy the current constraint, prohibiting setting $B_\infty = 0$ under the  constructibility analysis in Sec.~\ref{sec:constructibility} using the ALT shift. The factorized part of the amplitude in~Eq.~\eqref{eq:non-minimal-four-point} is nevertheless independent of $q_i$
and is little group covariant on its own. 
The violation of the current constraint generically results in higher mass divergences due to the numerator of the propagator and we expect $1/m^6$ divergence for non-minimal spin-3/2 Compton amplitude, but curiously,  
the above amplitude still has the same $1/m^4$ divergence as in the minimal case. One possibility is that the higher mass divergences reside in the boundary terms $B_\infty$ that was not accounted for here. These boundary terms may further characterize the amplitude’s dependence on off-shell interactions and contact terms in the spin-3/2 Lagrangian.

%%%%%%%%%%

%
\subsubsection{Spin-2 amplitude}
%%%%%%%%%%%%%%%%%%%%

Finally, we consider the $s=2$ four-point amplitude with the minimal three-point amplitude.
The motivation is similar to the investigation of the non-minimal three-point amplitude; the large-$z$ argument
does not guarantee $B_\infty = 0$, and we are curious to see the $q_i$ (in)dependence of the factorized part.

We start from Eqs.~\eqref{eq:t-channel_photon} and~\eqref{eq:u-channel_photon} with $s=2$,
%%%
\begin{align}
 \sum_{\lambda} \hat{A}_{\psi_{2} \bar{\psi}_{2}\gamma}^{(\lambda_1 \lambda -)}
	\times \hat{A}_{\psi_{2} \bar{\psi}_{2}\gamma}^{(\bar{\lambda} \lambda_2 +)}\bigg|_{z^{\pm}_{13}} 
    &=-\frac{([4|\hat{p}_1 | 3\rangle)^2([\textbf{1}|\hat{p}_{13}|\textbf{2}\rangle)^{4}}{\hat{t}_{14}m^{8}} \ , \\
 \sum_{\lambda} \hat{A}_{\psi_{2} \bar{\psi}_{2}\gamma}^{(\lambda_1 \lambda +)}
	\times \hat{A}_{\psi_{2} \bar{\psi}_{2}\gamma}^{(\bar{\lambda} \lambda_2 -)}\bigg|_{z^{\pm}_{14}}   &=-\frac{([4|\hat{p}_1 | 3\rangle])^2(\langle \textbf{1}|\hat{p}_{14}|\textbf{2}])^{4}}{\hat{t}_{13}m^{8}} \ .
\end{align}
%%%%%%%%%%
After some manipulation, we can rewrite them as 

\begin{equation}
\begin{aligned}
   \sum_{\lambda} \hat{A}_{\psi_{2} \bar{\psi}_{2}\gamma}^{(\lambda_1 \lambda -)}
	\times \hat{A}_{\psi_{2} \bar{\psi}_{2}\gamma}^{(\bar{\lambda} \lambda_2 +)}\bigg|_{z^{\pm}_{13}} 
    &=  -\frac{{\mathcal{T}}^2\hat{\mathcal{S}}^2  }{4\hat{t}_{14}m^4}  
    + \frac{1}{4m^6} {\mathcal{T}} {\mathcal{U}} (3\hat{\mathcal{X}}^2 +\hat{\mathcal{X}}\hat{\mathcal{Y}}) \ ,  \\
  \sum_{\lambda} \hat{A}_{\psi_{2} \bar{\psi}_{2}\gamma}^{(\lambda_1 \lambda +)}
	\times \hat{A}_{\psi_{2} \bar{\psi}_{2}\gamma}^{(\bar{\lambda} \lambda_2 -)}\bigg|_{z^{\pm}_{14}}
    &=  -\frac{{\mathcal{T}}^2\hat{\mathcal{S}}^2  }{4\hat{t}_{13}m^4}  
    + \frac{1}{4m^6} {\mathcal{T}} {\mathcal{V}} (3\hat{\mathcal{Y}}^2 + \hat{\mathcal{X}}\hat{\mathcal{Y}})  \ .
 \end{aligned}
 \end{equation}
%%%%%%%%%%%%%%%%%%%%%%%%%
The first terms are in the form such that we can use the complex analysis technique as before,
which reduces it to the residue of the pole at $z = 0$.
The second terms now contain up to quadratic powers in $z$, and the $z^2$ terms provide the contact term.
Indeed, by using
\begin{align}
	\frac{z_{1i}^+ (z_{1i}^-)^2 - z_{1i}^- (z_{1i}^+)^2}{z_{1i}^+ - z_{1i}^-}
	= -z_{1i}^+ z_{1i}^-
	= -\frac{p_{1i}^2 - m_I^2}{2q_1 \cdot q_i},
	\label{eq:z_square}
\end{align}
we can write the contact term as
\begin{align}
	\left.{A}_{\psi_{2}\bar{\psi}_{2}\gamma\gamma}^{(\lambda_1\lambda_2-+)}\right\vert_{\mathrm{cont.}}
	&= -\frac{\mathcal{T}}{4m^6}\left[\frac{\mathcal{U}}{2q_1 \cdot q_3}\left(3\mathcal{X}_q^2 +\mathcal{X}_q\mathcal{Y}_q\right)
	+ \frac{\mathcal{V}}{2q_1 \cdot q_4}\left(3\mathcal{Y}_q^2 +\mathcal{X}_q\mathcal{Y}_q\right)
	\right]
	+ B_\infty,
	\label{eq:spin2}
\end{align}
where the subscript ``$q$'' indicates that $p_i$ is replaced by $q_i$ in $\mathcal{X}$ and $\mathcal{Y}$.
These do not have the propagators anymore, providing the contact terms.
Since $q_i \propto \epsilon_i$ possesses the little group indices,
we should be able to eliminate the $q_i$ dependence if the amplitude is constructible under the ALT shift.\footnote{
	This is possible depending on the theory.
	For instance, this is possible for massive gauge boson scatterings in the electroweak theory,
	providing the gauge four-point contact interactions in the language of the Feynman diagrams~\cite{Ema:2024rss}.
} 
However, we could not find such a simplification of the above expression for $B_\infty=0$.
We have numerically checked that different choices of $q_i$ give different results, 
so the residual $q_i$ dependence is not an artifact due to insufficient simplification.
Since the original amplitude is little-group covariant, the residual $q_i$ dependence needs to be cancelled by $B_\infty \neq 0$.
Therefore, the spin-2 electromagnetic Compton amplitude in the minimal case is not on-shell constructible under the ALT shift.
This $q_i$ dependence, resulting in non-constructibility, resides in the term of order $1/m^6$.

%%%%%%%%%%%%%%%%%%%%%%%%%%%%
\subsubsection*{General spin-$s$ Compton amplitude}
For a general spin-$s$ four-point Compton amplitude, the $t$-channel contribution can be written as
%%%%%%%%%%%%%
\begin{align}
    \sum_{\lambda} \hat{A}_{\psi_{s} \bar{\psi}_{s}\gamma}^{(\lambda_1 \lambda -)}
	\times \hat{A}_{\psi_{s} \bar{\psi}_{s}\gamma}^{(\bar{\lambda} \lambda_2 +)}\bigg|_{z^{\pm}_{13}} 
     &= \tilde{\hat{x}}_{1I} {\hat{x}}_{I2}   \frac{[\textbf{1} \hat{\textbf{I}} ]^{2s}}{m^{4s-2}}  \left[ {[-\hat{\textbf{I}} \textbf{2} ]}- \frac{\hat{\tilde{x}}_{I2}} {m}   {[-\hat{\textbf{I}} 4 ][ 4 \textbf{2}]} \right]^{2s}\nonumber \\
    &= (-)^{2s+1}\frac{[4\lvert \hat{p}_1 \lvert 3 \rangle^2 }{\hat{t}_{14}m^{2s}}
    \left[{[\textbf{1} \textbf{2} ]^{2s}}- 2s \hat{\tilde{x}}_{I2} \frac{[\textbf{1}4 ][ 4 \textbf{2}][\textbf{1} \textbf{2} ]^{2s-1}}{m} + \begin{pmatrix}
        2s\\
        2
\end{pmatrix}\hat{\tilde{x}}_{I2}^2\frac{([\textbf{1}4 ][ 4 \textbf{2}])^2[\textbf{1} \textbf{2} ]^{2s-2}}{m^{2}} 
    \right. \nonumber \\ 
     &~~~~~~~~~\left.
     - \begin{pmatrix}
        2s\\
        3
\end{pmatrix} \hat{\tilde{x}}_{I2}^3 \frac{([\textbf{1}4 ][ 4 \textbf{2}])^3[\textbf{1} \textbf{2} ]^{2s-3}}{m^{3}}+ \begin{pmatrix}
        2s\\
        4
\end{pmatrix}\hat{\tilde{x}}_{I2}^4 \frac{([\textbf{1}4 ][ 4 \textbf{2}])^4[\textbf{1} \textbf{2} ]^{2s-4}}{m^{4}}- ... \right], 
     \label{eq:general_spin}
\end{align}
%%%%%%%%%%%%%%%%%%%%%
where $\hat{p}_I= - \hat{p}_{13}$.
Each term in the above expansion involves an increasing insertion of the spin operator $S$~\cite{Guevara:2018wpp}. The above series terminates at $\mathcal{O}(S^{2s})$ and we have explicitly verified this expansion up to  $\mathcal{O}(S^4)$ order with $s=2$. 
Eq.~\eqref{eq:general_spin} highlights the universal structure of the factorized amplitude at each order in spin coupling. Due to on-shell constructibility up to $s \leq 3/2$ and the universal structure of the factorized amplitude, 
the terms up to $\mathcal{O}(S^3)$ do not introduce any $q_i$ dependence. 
However, the terms of $\mathcal{O}(S^4)$ and higher have residual $q_i$ dependences.
Each increment in spin coupling introduces an additional factor of $1/m$ and an extra factor of $z^2$ in the numerator. 
For the spin-2 Compton amplitude we studied above, 
the residual $q_i$ dependence can be traced back to these additional $z$-factors in the numerator.

%%%%%%%%%%%%%%%%%%%%%%%%%%%%%%%
\section{Graviton Compton amplitude}\label{sec:comptongraviton}
%%%%%%%%%%%%%%%%%%%%%%%%%%%%%%%

In this section, we study the four-point gravitational Compton amplitudes by the ALT shift.
As we argued in Sec.~\ref{sec:constructibility}, the amplitude behaves as\footnote{Recall that we have also constrained the kinematic factors such that the propagator has no mass divergence, and the interactions are up to dimension 6.}
\begin{align}
	\lim_{\vert z \vert \to \infty}\hat{\mathcal{A}}_4 \sim z^{\gamma}, \quad \gamma \leq  - [g] - \frac{N_F}{2} -4,
\end{align}
where ``$-4$'' comes from the graviton polarization tensors. 
Restricting ourselves to the minimal three-point amplitudes, we have $[g] \geq -4$   for $s \leq 5/2$, and therefore the graviton
four-point amplitudes are on-shell constructible for $s \leq 5/2$. We can extend the analysis to higher-point Compton scatterings, 
for which the large-$z$ behavior is further improved. 
For $s>5/2$, the dimensional analysis does not guarantee the constructibility of the four-point Compton amplitude, and higher spin-$s$ amplitude can have additional undetermined contact terms beyond contributions from the factorized amplitudes. In the following, we calculate the graviton Compton four-point amplitudes explicitly.

%%%%%%%%%%%%%%%%%%%%%%%%%%%%%%%
\subsection{Four-point Compton amplitude}
%%%%%%%%%%%%%%%%%%%%%%%%%%%%%%%
We label the massive particles as $\textbf{1}$ and $\textbf{2}$ and the gravitons as 3 and 4, respectively.
The amplitude has an $s$-channel pole due to the graviton self-interaction in addition to the $t$- and $u$-channel poles.
They are located at
\begin{align}
	\hat{p}_{13}^2 = m^2,
	\quad
	\hat{p}_{14}^2 = m^2,
	\quad
	\hat{p}_{12}^2 = 0.
\end{align}
As a result, we have
\begin{align}
	{A}_{\psi_s\bar{\psi}_shh}^{(\lambda_1\lambda_2\lambda_3\lambda_4)}
	&= \sum_{i=3,4}\frac{1}{p_{1i}^2 - m^2}\frac{1}{z_{1i}^+ - z_{1i}^-}
	\sum_\lambda
	\left[z_{1i}^+ \hat{A}_{\psi_s \bar{\psi}_sh}^{(\lambda_1 \lambda \lambda_i)}(z_{1i}^-)
	\times \hat{A}_{\psi_s \bar{\psi}_sh}^{(\bar{\lambda} \lambda_2 \lambda_j)}(z_{1i}^-)
	- (z_{1i}^+ \leftrightarrow z_{1i}^-)
	\right]
	\nonumber \\
	&+\frac{1}{p_{12}^2}\frac{1}{z_{12}^+ - z_{12}^-}
	\sum_\lambda
	\left[
	z_{12}^+ \hat{A}_{\psi_s \bar{\psi}_sh}^{(\lambda_1 \lambda_2 \lambda)}(z_{12}^-)
	\times \hat{A}_{hhh}^{(\bar{\lambda} \lambda_3 \lambda_4)}(z_{12}^-)
	- (z_{12}^+ \leftrightarrow z_{12}^-)
	\right]
	+ B_\infty,
    \label{eq:sum_graviton}
\end{align}
where $j \neq i, 1,2$ and $\bar{\lambda}$ is the little group conjugate of $\lambda$. Again, $z_{1i}^{\pm
}$ are the two solutions of $\hat{t}_{1i}= 2 \hat{p}_1 \cdot \hat{p}_i=0$ for $i=3,4$, and $z_{12}^{\pm}$ are the two solutions of $\hat{p}_{12}^2=\hat{t}_{34}=0$. 
The graviton three-point amplitude is completely fixed by the $U(1)$ little group and is given by\footnote{
%%%%%%
The relative size between $A_{hhh}$ and $A_{\psi_s \bar{\psi}_sh}$ can be fixed by requiring the factorized amplitude to take the form in Eq.~\eqref{eq:consistant}, which we verify explicitly up to $s=3$. 
%%%%%
}
\begin{align}
	A_{hhh}^{(++-)} = \frac{1}{M_P}\frac{[12]^6}{[23]^2[31]^2},
	\quad
	A_{hhh}^{(-+-)} = \frac{1}{M_P}\frac{\langle31\rangle^6}{\langle 12\rangle^2\langle 23\rangle^2}\ .
\end{align}

%%%%%%%%%%%%
\subsubsection*{Same-helicity case}
%%%%%%%%%%%%

In the case of the same helicity gravitons,  the calculation is similar to the photon case, apart from the additional $s$-channel contribution. 
For $\lambda_3 = \lambda_4 = +$, the amplitude, after summing over all the poles, is given by  
\begin{align}  
{A}_{\psi_s\bar{\psi}_shh}^{(\lambda_1\lambda_2++)} &= (-)^{2s+1}\frac{\langle \mathbf{12}\rangle^{2s} [34]^4}{ M_P^2m^{2s-4}t_{14}{t}_{13}{t}_{34}}
+ B_\infty.  
\end{align}  
Again, the factorized part is little group covariant for general $s$ even though setting $B_\infty = 0$ is justified only for $s \leq 5/2$,
and this part reproduces the result in~\cite{Aoude:2020onz}.

%%%%%%%%%%%%
\subsubsection*{Opposite-helicity case}
%%%%%%%%%%%%
For $\lambda_3 = -$ and $\lambda_4 = +$,
the products of the minimal three-point amplitudes for the $t$- and $u$-channels are given by
\begin{align}
	\sum_{\lambda} \hat{A}_{\phi_s \bar{\phi}_sh}^{(\lambda_1 \lambda -)}(z_{13}^\pm)
	\times \hat{A}_{\phi_s \bar{\phi}_sh}^{(\bar{\lambda} \lambda_2 +)}(z_{13}^\pm) 
	&= - \frac{([4|\hat{p}_1 | 3\rangle)^4([\textbf{1}]|\hat{p}_{13}|\textbf{2}\rangle)^{2s}}{\hat{t}_{14}\hat{t}_{34} M_P^2 m^{4s-2}}
   \ ,  \label{eq:t-channel-grav}
\end{align}
and
\begin{align}
	\sum_{\lambda} \hat{A}_{\phi_s \bar{\phi}_sh}^{(\lambda_1 \lambda +)}(z_{14}^\pm)
	\times \hat{A}_{\phi_s \bar{\phi}_sh}^{(\bar{\lambda} \lambda_2 -)}(z_{14}^\pm)
	&= - \frac{([4|\hat{p}_1 | 3\rangle)^4(\langle \textbf{1}]|\hat{p}_{14}|\textbf{2}])^{2s}}{\hat{t}_{13}\hat{t}_{34} M_P^2 m^{4s-2}},
    \label{eq:u-channel-grav}
\end{align}
while for the $s$-channel, we obtain
\begin{align}
	\sum_\lambda \hat{A}_{\phi_s \bar{\phi}_s h}^{(\lambda_1\lambda_2 \lambda)} (z_{12}^{\pm})\times \hat{A}_{hhh}^{(\bar{\lambda}-+)} (z_{12}^{\pm})
	&=(-)^{2s+1}
	\frac{\left(\langle \mathbf{1}3\rangle [\mathbf{2}4] + [ \mathbf{1}4] 
    \langle \mathbf{2}3 \rangle \right)^{2s}}
	{\hat{t}_{13} \hat{t}_{14}M_P^2 }
	\frac{1}{[ 4\vert \hat{p}_{1}\vert 3\rangle^{2s-4}} \ .
    \label{eq:s-channel_graviton}
\end{align}
As in the photon case, after using the Schouten identity and the pole condition for the $t$- and $u$-channels, 
the factorized part can be rewritten as
\begin{align}
	{A}_{\psi_{s}\bar{\psi}_{s}hh}^{(\lambda_1\lambda_2-+)}
	&= (-)^{2s+1} \frac{\left(\langle \mathbf{1}3\rangle [\mathbf{2}4] + [ \mathbf{1}4] 
    \langle \mathbf{2}3 \rangle \right)^{2s}}{M_P^2}
	\sum \mathrm{Res}_{z=z_{12}^\pm,z_{13}^\pm,z_{14}^\pm}
	\left[-\frac{1}{z}\frac{1}{\hat{t}_{34}\hat{t}_{13}\hat{t}_{14}}\frac{1}{[ 4\vert \hat{p}_1 \vert 3\rangle^{2s-4}}
	\right]
	+ B_\infty.
\end{align}
The term in the square bracket vanishes sufficiently fast at $\vert z \vert \to \infty$, and therefore we obtain the amplitude
%%%
\begin{align}
	{A}_{\psi_{s}\bar{\psi}_{s}hh}^{(\lambda_1\lambda_2-+)}
	&= (-)^{2s+1}\frac{\left(\langle \mathbf{1}3\rangle [\mathbf{2}4] + [ \mathbf{1}4] 
    \langle \mathbf{2}3 \rangle 
 \right)^{2s}}{M_P^2}
	\frac{[ 4\vert {p}_1 \vert 3\rangle^{4-2s}}{t_{34}t_{13}t_{14}},
\end{align}
for $0 \leq s \leq 2$ where we set $B_\infty = 0$, and for $s \geq 5/2$
\begin{align}
	{A}_{\psi_{s}\bar{\psi}_{s}hh}^{(\lambda_1\lambda_2-+)}
	&= (-)^{2s+1}\frac{\left(\langle \mathbf{1}3\rangle [\mathbf{2}4] + [ \mathbf{1}4] 
    \langle \mathbf{2}3 \rangle 
 \right)^{2s}}
	{M_P^2 [ 4\vert {p}_1 \vert 3\rangle^{2s-4}}
	\left[
	\frac{1}{t_{34}t_{13}t_{14}}+
	\frac{(-z_\mathrm{sp})^{2s-4}}{(2s-5)!}
	\frac{d^{2s-5}}{dz^{2s-5}}\left(\frac{1}{z}\frac{1}{\hat{t}_{34}\hat{t}_{13}\hat{t}_{14}}\right)_{z\to z_\mathrm{sp}}
	\right]
	+ B_\infty.
    \label{eq: sp_pole_subtract}
\end{align}
In the following, we focus on $s = 5/2$ and simplify this expression.

%%%%%%%%%%
\subsubsection {Spin-5/2 amplitude}
%%%%%%%%%%

By setting $s = 5/2$ in Eq.~\eqref{eq: sp_pole_subtract}, we obtain
\begin{align}
	{A}_{\psi_{5/2}\bar{\psi}_{5/2}hh}^{(\lambda_1\lambda_2-+)}
	&= \frac{\left(\langle \mathbf{1}3\rangle [\mathbf{2}4] + [ \mathbf{1}4] 
    \langle \mathbf{2}3 \rangle 
 \right)^{5}}{M_P^2
	[ 4\vert p_1 \vert 3 \rangle}
	\left[\frac{1}{t_{34}t_{13}t_{14}}
	- \left.\frac{1}{\hat{t}_{34}\hat{t}_{13}\hat{t}_{14}}\right\vert_{z = z_\mathrm{sp}}
	\right].
\end{align}
By noting that, at $z = z_\mathrm{sp}$ where $[ 4\vert \hat{p}_1 \vert 3\rangle = 0$,
\begin{align}
	\hat{t}_{13}\hat{t}_{14}
	= \langle \hat{3}\vert \hat{p}_1 \vert 3] \langle 4 \vert \hat{p}_1 \vert \hat{4}]
	= -\langle \hat{3} 4\rangle [3 \vert \hat{p}_1^2 \vert \hat{4}]
	= m^2 \hat{t}_{34},
\end{align}
together with Eq.~\eqref{eq:sp_pole_identity}, we can simplify the second term and obtain
\begin{align}
	{A}_{\psi_{5/2}\bar{\psi}_{5/2}hh}^{(\lambda_1\lambda_2-+)}
    &= \frac{\langle \mathbf{1}3\rangle [\mathbf{2}4] + [ \mathbf{1}4] 
    \langle \mathbf{2}3 \rangle }{M_P^2
	[ 4\vert p_1 \vert 3\rangle }
	\left[\frac{\left(\langle \mathbf{1}3\rangle [\mathbf{2}4] + [ \mathbf{1}4] 
    \langle \mathbf{2}3 \rangle 
 \right)^{4}}
	{t_{34}t_{13}t_{14}}
	- \frac{[\mathbf{1}4]^2 \langle\mathbf{2}3\rangle^2\langle \mathbf{1}3\rangle^2 [ \mathbf{2}4]^2}{m^6}
	\right].
    \label{eq:graviton-Compton1}
\end{align}
The result is independent of $q_i$ and is free from a spurious pole since the spurious pole 
at $z = z_\mathrm{sp}$, where $[ 4\vert p_1 \vert 3\rangle = 0$, contained in the first term is cancelled by the second term.
Again, we emphasize that the second term is uniquely fixed in our procedure with the explicit momentum shift,
as opposed to ad hoc spurious pole subtraction by hand without a specific momentum shift.

As in the photon case, we can make our result explicitly free from the spurious pole.
For this purpose, we may go back to Eqs.~\eqref{eq:t-channel-grav}\,--\,\eqref{eq:s-channel_graviton}
and bring them to the form
\begin{equation}
\begin{aligned}
 \sum_{\lambda} \hat{A}_{\psi_s \bar{\psi}_s\gamma}^{(\lambda_1 \lambda -)}(z_{13}^\pm)
	\times \hat{A}_{\psi_s \bar{\psi}_s\gamma}^{(\bar{\lambda} \lambda_2 +)}(z_{13}^\pm)
	 &=\frac{\hat{f}_{stu}}{\hat{t}_{14}\hat{t}_{34}} + \frac{\hat{f}_{tu}}{\hat{t}_{14}}+\frac{\hat{f}_{ts}}{\hat{t}_{34}}+\hat{f}_t \ ,\\ 
      \sum_{\lambda} \hat{A}_{\psi_s \bar{\psi}_s\gamma}^{(\lambda_1 \lambda +)}(z_{14}^\pm)
	\times \hat{A}_{\psi_s \bar{\psi}_s\gamma}^{(\bar{\lambda} \lambda_2 -)}(z_{14}^\pm)
	 &=\frac{\hat{f}_{stu}}{\hat{t}_{13}\hat{t}_{34}} + \frac{\hat{f}_{tu}}{\hat{t}_{13}}+\frac{\hat{f}_{us}}{\hat{t}_{34}}+\hat{f}_u  \ ,  \\
  \sum_\lambda \hat{A}_{\psi_s \bar{\psi}_s h}^{(\lambda_1\lambda_2 \lambda)} (z_{12}^{\pm})\times \hat{A}_{hhh}^{(\bar{\lambda}-+)} (z_{12}^{\pm})
	&=\frac{\hat{f}_{stu}}{\hat{t}_{13}\hat{t}_{14}} + \frac{\hat{f}_{ts}}{\hat{t}_{13}}+\frac{\hat{f}_{us}}{\hat{t}_{14}}+\hat{f}_s  \ . 
\end{aligned}
\label{eq:consistant}
\end{equation}
%%%%
We can express Eq.~\eqref{eq:t-channel-grav} as
\begin{align}
	\sum_{\lambda} \hat{A}_{\psi_{5/2} \bar{\psi}_{5/2}h}^{(\lambda_1 \lambda -)}(z_{13}^\pm)
	\times \hat{A}_{\psi_{5/2} \bar{\psi}_{5/2}h}^{(\bar{\lambda} \lambda_2 +)}(z_{13}^\pm)
	&= -\frac{([4|\hat{p}_1 | 3\rangle)^4([\textbf{1}]|\hat{p}_{13}|\textbf{2}\rangle)^{5}}{\hat{t}_{14}\hat{t}_{34} M_P^2 m^{8}}
    =  -\frac{([4|\hat{p}_1 | 3\rangle)^4(\hat{\mathcal{S}} - [\textbf{1}\lvert \frac{\hat{p}_3 \hat{p}_4}{m}\lvert \textbf{2}])^{5}}{2^5\hat{t}_{14}\hat{t}_{34} M_P^2 m^{8}}  .
    \label{eq:t-channel-grav1}
\end{align}
By employing Eq.~\eqref{eq:id} after expanding the terms in the bracket and Eq.~\eqref{eq:id2} to minimize powers of $\hat{t}_{14}$, 
we reduce it to
\begin{align}
  \sum_{\lambda} \hat{A}_{\psi_{5/2} \bar{\psi}_{5/2}h}^{(\lambda_1 \lambda -)}(z_{13}^\pm)
	\times \hat{A}_{\psi_{5/2} \bar{\psi}_{5/2}h}^{(\bar{\lambda} \lambda_2 +)}(z_{13}^\pm) &= -\frac{1}{M_P^2}\bigg(\frac{\hat{\mathcal{S}}{\mathcal{T}}^4}{2m^2\hat{t}_{14}\hat{t}_{34}} - \frac{\hat{\mathcal{S}}{\mathcal{T}}^3{\mathcal{U}}}{4m^4\hat{t}_{34}}-\frac{{\mathcal{T}}^2{\mathcal{U}}^2\hat{\mathcal{X}}}{4m^6}\bigg) .
\end{align}
Similarly, we obtain for $u$-channel ($\hat{t}_{14}=0$) 
%%%%
\begin{align}
 \sum_{\lambda} \hat{A}_{\psi_{5/2} \bar{\psi}_{5/2}h}^{(\lambda_1 \lambda +)}(z_{14}^\pm)
	\times \hat{A}_{\psi_{5/2} \bar{\psi}_{5/2}h}^{(\bar{\lambda} \lambda_2 -)}(z_{14}^\pm) &= -\frac{1}{M_P^2}\bigg(\frac{\hat{\mathcal{S}}{\mathcal{T}}^4}{2m^2\hat{t}_{13}\hat{t}_{34}} - \frac{\hat{\mathcal{S}}{\mathcal{T}}^3{\mathcal{V}}}{4m^4\hat{t}_{34}}-\frac{ {\mathcal{T}}^2{\mathcal{V}}^2\hat{\mathcal{Y}}}{4m^6}\bigg).
\end{align}
%%%%%
For $s$-channel ($\hat{t}_{34}=0$), we rewrite Eq.~\eqref{eq:s-channel_graviton} for $s=5/2$ with the following relations

%%%%%%%
\begin{align}
    \frac{{\mathcal{T}}}{[4\lvert \hat{p}_1 \lvert 3 \rangle}  &= - \frac{\hat{\mathcal{S}}}{2m^2} - \frac{\hat{t}_{13} \mathcal{V}}{2 m^2 [4\lvert \hat{p}_1 \lvert 3 \rangle} - \frac{\hat{t}_{14} \mathcal{U}}{2m^2 [4\lvert \hat{p}_1 \lvert 3 \rangle}  = \frac{\hat{ \mathcal{Y}}- \hat{\mathcal{X}} }{\hat{t}_{13}}\Bigg|_{z=z_{12}^{\pm}},
\end{align}
%%%%%%%
and obtain
%%%%%
\begin{align}
  \sum_\lambda \hat{A}_{\psi_{5/2} \bar{\psi}_{5/2} h}^{(\lambda_1\lambda_2 \lambda)} (z_{12}^{\pm})\times \hat{A}_{hhh}^{(\bar{\lambda}+-)} (z_{12}^{\pm}) &= -\frac{1}{M_P^2}\left[\frac{\hat{\mathcal{S}}{\mathcal{T}}^4}{2m^2\hat{t}_{13}\hat{t}_{14}} - \frac{\hat{\mathcal{S}}{\mathcal{T}}^3{\mathcal{U}}}{4m^4\hat{t}_{13}}- \frac{\hat{\mathcal{S}}{\mathcal{T}}^3{\mathcal{V}}}{4m^4\hat{t}_{14}} -\frac{{\mathcal{T}} ({\mathcal{U}}-{\mathcal{V}})^2(\hat{\mathcal{Y}}{\mathcal{V}}+\hat{\mathcal{X}} {\mathcal{U}})}{4m^6}\right]. 
\end{align}
%%%%%
After summing over all the poles in Eq.~\eqref{eq:sum_graviton}, we then derive the spin-5/2 graviton Compton amplitude as
%%%%%%
\begin{tcolorbox}[colback=gray!5, colframe=black, boxrule=0.2mm]
\begin{align}
    &{A}_{\psi_{5/2}\bar{\psi}_{5/2}hh}^{(\lambda_1\lambda_2-+)} 
    \nonumber \\
    &= - \frac{1}{M_P^2} \bigg[
        \frac{\mathcal{S} \mathcal{T}^4}{2m^2 t_{13} t_{14} t_{34}} 
        -\frac{\mathcal{S}\mathcal{T}^3}{4m^4t_{34}}\left[\frac{\mathcal{U}}{t_{13}}+\frac{\mathcal{V}}{t_{14}}\right]
    - \frac{\mathcal{T}^2}{4m^6}\left[\frac{\mathcal{U}^2 \mathcal{X}}{t_{13}}+\frac{\mathcal{V}^2 \mathcal{Y}}{t_{14}}\right]
        - \frac{\mathcal{T} (\mathcal{U} - \mathcal{V})^2 (\mathcal{Y} \mathcal{V} + \mathcal{X} \mathcal{U})}{4m^6 t_{34}}
    \bigg] \nonumber \\
    &= - \frac{1}{M_P^2} \left[
    \big( [\mathbf{2}4] \langle \mathbf{1}3 \rangle 
        + [\mathbf{1}4] \langle \mathbf{2}3 \rangle \big)^3 
        \frac{[\mathbf{1} | p_{13} | \mathbf{2} \rangle 
        + \langle \mathbf{1} | p_{14}| \mathbf{2} ]}{2m^2 t_{34}}
     \left[
        \frac{[\mathbf{2}4] \langle \mathbf{1}3 \rangle + [\mathbf{1}4] \langle \mathbf{2}3 \rangle}{t_{13} t_{14} }
        - \frac{[\mathbf{2}4] \langle \mathbf{1}3 \rangle}{2m^2 t_{13} } 
        - \frac{[\mathbf{1}4] \langle \mathbf{2}3 \rangle}{2m^2 t_{14} } 
    \right] \right. \nonumber \\
    &~~~~~~~~~~~~ - \big( [\mathbf{2}4] \langle \mathbf{1}3 \rangle 
        + [\mathbf{1}4] \langle \mathbf{2}3 \rangle \big)^2 \left(
        \frac{[\mathbf{2}4]^2 \langle \mathbf{1}3 \rangle^2 [\mathbf{1} | p_{13} | \mathbf{2} \rangle}{4m^6 t_{13}}
        + \frac{[\mathbf{1}4]^2 \langle \mathbf{2}3 \rangle^2 \langle \mathbf{1} | p_{14} | \mathbf{2} ]}{4m^6 t_{14}}
    \right) \nonumber \\
    &~~~~~~~~~~~~ \left.- \frac{
        \big( [\mathbf{2}4] \langle \mathbf{1}3 \rangle + [\mathbf{1}4] \langle \mathbf{2}3 \rangle \big)
        \big( [\mathbf{2}4] \langle \mathbf{1}3 \rangle - [\mathbf{1}4] \langle \mathbf{2}3 \rangle \big)^2
        \big( [\mathbf{1} | p_{13} | \mathbf{2} \rangle [\mathbf{2}4] \langle \mathbf{1}3 \rangle 
        + \langle \mathbf{1} | p_{14} | \mathbf{2} ] [\mathbf{1}4] \langle \mathbf{2}3 \rangle \big)
    }{4m^6 t_{34}} 
    \right]  \ .
    \label{eq:graviton-Compton}
\end{align}
\end{tcolorbox}
\noindent
One can check that Eqs.~\eqref{eq:graviton-Compton1} and~\eqref{eq:graviton-Compton} are equivalent. 
This result is explicitly free from spurious poles and is in agreement with~\cite{Chiodaroli:2021eug}. 
We note that the amplitude exhibits a mass divergence of $1/m^6$. 
Table~\ref{table:graviton} summarizes the high-energy behavior of the Compton amplitude in the CM frame. The pattern of high-energy behavior for spin-5/2 graviton Compton is similar to spin-3/2 photon Compton amplitude. In the case of same helicity graviton ($h_3,h_4=+,+$), the leading energy behavior of $E^3$ comes from the MHV helicity configuration $(-\frac{5}{2},-\frac{5}{2})$, with the remaining entries in the table showing energy suppression due to helicity flips. For the case of opposite graviton helicity ($h_3,h_4=-,+$),  the high-energy behavior in the top-right 4 $\times$ 4 block of the table is determined by the second term in the bracket of Eq.~\eqref{eq:graviton-Compton1}, while the remaining entries are determined by the first term. 

%%%%%%%%%%%
\begin{table}[h]
\centering
\renewcommand{\arraystretch}{1.4}
\large

\begin{minipage}{0.45\textwidth}
\centering
\scalebox{0.93}{
\begin{tabular}{|c|c|c|c|c|c|c|}
\hline
\multicolumn{7}{|c|}{\((h_3, h_4) = (-,+)\)} \\
\hline
\diagbox{$\lambda_2$}{$\lambda_1$} & $+\frac{5}{2}$ & $+\frac{3}{2}$ & $+\frac{1}{2}$ & $-\frac{1}{2}$ & $-\frac{3}{2}$ & $-\frac{5}{2}$ \\
\hline
$+\frac{5}{2}$ & $E^{-3}$ & $E^{-2}$ & $E^{5}$ & $E^{6}$ & $E^{5}$ & $E^{4}$ \\
\hline
$+\frac{3}{2}$ & $E^{-2}$ & $E^{-1}$ & $E^{6}$ & $E^{7}$ & $E^{6}$ & $E^{5}$ \\
\hline
$+\frac{1}{2}$ & $E^{-1}$ & $E^{0}$ & $E^{7}$ & $E^{8}$ & $E^{7}$ & $E^{6}$ \\
\hline
$-\frac{1}{2}$ & $E^{0}$ & $E^{1}$ & $E^{8}$ & $E^{7}$ & $E^{6}$ & $E^{5}$ \\
\hline
$-\frac{3}{2}$ & $E^{1}$ & $E^{2}$ & $E^{1}$ & $E^{0}$ & $E^{-1}$ & $E^{-2}$ \\
\hline
$-\frac{5}{2}$ & $E^{2}$ & $E^{1}$ & $E^{0}$ & $E^{-1}$ & $E^{-2}$ & $E^{-3}$ \\
\hline
\end{tabular}
}
\end{minipage}
\hspace{1cm}
\begin{minipage}{0.45\textwidth}
\centering
\scalebox{0.93}{
\begin{tabular}{|c|c|c|c|c|c|c|}
\hline
\multicolumn{7}{|c|}{\((h_3, h_4) = (+,+)\)} \\
\hline
\diagbox{$\lambda_2$}{$\lambda_1$} & $+\frac{5}{2}$ & $+\frac{3}{2}$ & $+\frac{1}{2}$ & $-\frac{1}{2}$ & $-\frac{3}{2}$ & $-\frac{5}{2}$ \\
\hline
$+\frac{5}{2}$ & $E^{-7}$ & $E^{-6}$ & $E^{-5}$ & $E^{-4}$ & $E^{-3}$ & $E^{-2}$ \\
\hline
$+\frac{3}{2}$ & $E^{-6}$ & $E^{-5}$ & $E^{-4}$ & $E^{-3}$ & $E^{-2}$ & $E^{-1}$ \\
\hline
$+\frac{1}{2}$ & $E^{-5}$ & $E^{-4}$ & $E^{-3}$ & $E^{-2}$ & $E^{-1}$ & $E^{0}$ \\
\hline
$-\frac{1}{2}$ & $E^{-4}$ & $E^{-3}$ & $E^{-2}$ & $E^{-1}$ & $E^{0}$ & $E^{1}$ \\
\hline
$-\frac{3}{2}$ & $E^{-3}$ & $E^{-2}$ & $E^{-1}$ & $E^{0}$ & $E^{1}$ & $E^{2}$ \\
\hline
$-\frac{5}{2}$ & $E^{-2}$ & $E^{-1}$ & $E^{0}$ & $E^{1}$ & $E^{2}$ & $E^{3}$ \\
\hline
\end{tabular}
}
\end{minipage}

\caption{Energy dependence of different helicity configurations for the spin-$\frac{5}{2}$ graviton Compton amplitude in the center-of-mass frame. Here, particles 1 and 3, and particles 2 and 4, are chosen to have anti-aligned spatial momentum. The variable $E$ denotes the CM energy.}
\label{table:graviton}
\end{table}

%%%%%%%%%%%%%%%%%%
\subsubsection {Spin-3 amplitude}
%%%%%%%%%%%%%%%%%%

As we have argued, setting $B_\infty = 0$ is justified up to $s \leq 5/2$ for the graviton Compton amplitudes.
Nevertheless, it would be instructive to see explicitly the property of the factorized part of the amplitude for $s > 5/2$.
For this purpose, here we briefly study the factorized part for $s = 3$.

For $s = 3$, the factorized parts of the opposite-helicity Compton amplitudes are given by
\begin{align}
  \sum_{\lambda} \hat{A}_{\psi_{3} \bar{\psi}_{3}h}^{(\lambda_1 \lambda -)}(z_{13}^\pm)
	\times \hat{A}_{\psi_{3} \bar{\psi}_{3}h}^{(\bar{\lambda} \lambda_2 +)}(z_{13}^\pm) 
	= - \frac{1}{M_P^2}
	&\left[\frac{\hat{\mathcal{S}}^2{\mathcal{T}}^4}{4m^4\hat{t}_{14}\hat{t}_{34}} 
	- \frac{{\mathcal{T}}^3{\mathcal{U}}\hat{\mathcal{X}}(3 \hat{\mathcal{X}}+\hat{\mathcal{Y}} )}{4m^6\hat{t}_{34}}\right], 
	\\
      \sum_{\lambda} \hat{A}_{\psi_{3} \bar{\psi}_{3}h}^{(\lambda_1 \lambda +)}(z_{14}^\pm)
	\times \hat{A}_{\psi_{3} \bar{\psi}_{3}h}^{(\bar{\lambda} \lambda_2 -)}(z_{14}^\pm) 
	= -\frac{1}{M_P^2}&\left[\frac{\hat{\mathcal{S}}^2{\mathcal{T}}^4}{4m^4\hat{t}_{13}\hat{t}_{34}} - \frac{{\mathcal{T}}^3{\mathcal{V}}\hat{\mathcal{Y}}(3 \hat{\mathcal{Y}}+\hat{\mathcal{X}} )}{4m^6\hat{t}_{34}}\right],
\end{align}
for the $t$- and $u$-channels, and
\begin{align}
         \sum_\lambda \hat{A}_{\psi_{3} \bar{\psi}_{3} h}^{(\lambda_1\lambda_2 \lambda)} (z_{12}^{\pm})\times \hat{A}_{hhh}^{(\bar{\lambda}+-)} (z_{12}^{\pm})  =- \frac{1}{M_P^2}&\left[\frac{\hat{\mathcal{S}}^2{\mathcal{T}}^4}{4m^4\hat{t}_{13}\hat{t}_{14}}- \frac{{\mathcal{T}}^3{\mathcal{U}}\hat{\mathcal{X}}(3 \hat{\mathcal{X}}+\hat{\mathcal{Y}} )}{4m^6\hat{t}_{13}}- \frac{{\mathcal{T}}^3{\mathcal{V}}\hat{\mathcal{Y}}(3 \hat{\mathcal{Y}}+\hat{\mathcal{X}} )}{4m^6\hat{t}_{14}} \right.
         \nonumber \\ 
         &\left. +\frac{\mathcal{U}\mathcal{V}(\mathcal{U}\hat{\mathcal{X}}+\mathcal{V}\hat{\mathcal{Y}})(2\mathcal{U}\hat{\mathcal{X}}+\mathcal{U}\hat{\mathcal{Y}}+\mathcal{V}\hat{\mathcal{X}}+2\mathcal{V}\hat{\mathcal{Y}})}{m^8}\right],
\label{eq:spin3}
\end{align}
for the $s$-channel, respectively.
We see that the amplitude can be written in the form Eq.~\eqref{eq:consistant}. 
In the $s$-channel, the term in the second line contains two momentum insertions without additional factors of $z$ in the denominator, and they generate contact terms after summing over all poles. The corresponding contact terms have residual $q_i$ dependence (we have checked this analytically and numerically).
These terms indicate that spin-3 and higher spin gravitational Compton amplitudes, with the minimal three-point amplitudes, are not on-shell constructible by the ALT shift. 
The constructibility breaks down at $\mathcal{O}(S^6)$. Since the complete amplitude is independent of $q_i$, it can be useful to check how the cancellation of $q_i$ occurs between the factorized amplitude and the terms of $B_\infty$, and whether this cancellation imposes additional constraints on the possible form of the contact terms for the full amplitude. 

%%%%%%%
%%%%%%%%%%%%
\section{Conclusion}\label{sec:conclusion}
A consistent interacting theory for higher-spin particles represents a longstanding question. For charged higher-spin particles, though some equations of motion exist~\cite{Benakli:2023aes}, writing down a decoupled form of the Lagrangian is particularly challenging. 
Despite the absence of a complete off-shell description, one can investigate certain physical observables using on-shell methods. 
In this work, we have employed the on-shell recursion relation to systematically calculate higher-spin Compton amplitudes.
With the minimal three-point amplitudes as the input, we have demonstrated that the electromagnetic and gravitational Compton amplitudes are constructible up to the spin $s \leq 3/2$ and $s \leq 5/2$, respectively, under the ALT momentum shift, and provide the full expressions in different equivalent forms.
The constructed four-point Compton amplitudes are manifestly little-group covariant and are \textit{free from spurious poles and contact term ambiguities}. Our main results are presented in Eqs.~\eqref{eq:four-point-compton-three-half} and~\eqref{eq:four-Compton-three-half} for the photon case, 
and in Eqs.~\eqref{eq:graviton-Compton1} and~\eqref{eq:graviton-Compton} for the graviton case, respectively. In this regard, we also prove the antsaz of spurious pole subtraction discussed in~\cite{Chiodaroli:2021eug} and show explicitly where it yields the full results (when constructible) and where it fails (when non-constructible).
Since the four-point Compton amplitudes are crucial not only in the quantum theory of higher spin particles but also, e.g., the classical dynamics of
spinning black holes, our result potentially sheds new light across a wide variety of fields. In particular, we prove that higher-point Compton amplitudes can be constructed by the ALT shift without the spurious poles and contact term ambiguities, which may provide an insight into higher-order corrections to Kerr black hole scatterings. We hope to come back to this point in a future publication.

The mass dimension of the couplings and the current constraint are essential for the on-shell constructibility under the ALT shift,
which restricts us to focus on the ``minimal'' three-point amplitudes, defined in Eq.~\eqref{eq:min}.
In the electromagnetic case, the minimal three-point spin-3/2 amplitude recovers a gyromagnetic ratio $g=2$. 
In the gravitational case, as is detailed in App.~\ref{app:threepointspin52},   
we notice that the minimal spin-5/2 amplitude requires couplings with at least two derivatives, which conflicts with the equivalence principle (in the sense that the coupling is given solely by $h_{\mu\nu}T^{\mu\nu}$). This was pointed out in~\cite{Porrati:1993in} 
and it is interesting to see how this arises from an on-shell perspective. In general, one may expect to extract useful information from on-shell three-point amplitudes, that hints towards consistent Lagrangian interactions~\cite{Cangemi:2022bew,Cangemi:2023ysz,Cangemi:2023bpe}.

We have studied the high-energy behaviors of the constructed four-point amplitudes, 
as shown in Tables~\ref{tab:helicity_energy_spin_three_half} and~\ref{table:graviton}.
Even though the minimal three-point amplitudes are expected to exhibit a better UV behavior compared to the non-minimal ones, 
certain helicity amplitudes still violate unitarity at a low cutoff. 
This shall not be surprising as higher spin amplitudes generically require an additional field content for a UV completion. A well-known example is the Polonyi model in supergravity, where the gravitino scattering is unitary up to Planck scale, 
after introducing an additional complex scalar. 
In this respect, the on-shell methods may provide a powerful tool to investigate candidates of higher spin UV completions
systematically from the bottom-up perspective. 
We have already reproduced in this way the Higgs boson in the electroweak theory~\cite{Ema:2024rss}  and Polonyi models in supergravity~\cite{Gherghetta:2024tob}.

For $s > 3/2$ (resp.~$5/2$), our dimensional analysis does not guarantee that the photon (resp.~graviton) Compton scatterings 
are constructible under the ALT shift. 
We have explicitly calculated the factorized parts of the electromagnetic spin-2 and gravitational spin-3 amplitudes,
and shown that the result depends on the shifted momenta, implying the loss of constructibility; see the discussions around 
Eqs.~\eqref{eq:spin2} and~\eqref{eq:spin3}.
It remains an open question whether one can recover the constructibility with the inclusion of other (lower-spin) degrees of freedom. 

Finally, another future direction is to investigate the constructibility in the presence of higher-dimensional non-minimal couplings. 
We have taken a first step by examining the spin-3/2 electromagnetic Compton amplitude with non-minimal three-point amplitudes. 
Although our dimensional analysis does not guarantee the constructibility, curiously, the factorized part of the amplitude, calculated by the ALT shift, 
is still free from spurious poles and is independent of the shifted momentum; see Eq.~\eqref{eq:non-minimal-four-point}.
We stress that the dimensional analysis, presented in this paper, 
gives only an upper bound of the large-$z$ behavior, and the true $z$-scaling may be better than expected. 
A more stringent bound on the large-$z$ behavior, if possible, would allow us to extend the on-shell construction to 
a larger class of theories and even higher spin particles. 

%
%%%%%%%%%%%%%%%%%%%%%%%%%%%%%%%%%%%%%%%%%%%%%%%%%%%%%%%%%%%%

%%%%%%%%%%%%%%%%%%%%%%%%%%%%%%%%%%%%%%%
\section*{Acknowledgements}
%%%%%%%%%%%%%%%%%%%%%%%%%%%%%%%%%%%%%%%

We acknowledge Andreas Helset and Paolo Pichini for helpful discussion and comments on our manuscript.
This work is supported in part by a DOE grant \#DE-SC0011842 and a Sloan Research Fellowship 
from the Alfred P. Sloan Foundation at the University of Minnesota.

%

%%%%%%%%%%%%%%%%%%%%%%%%%%%%%%%%%%%%%%%%%%%%%%%%%%%%%%%%%%%%
%%%%%%%%%%%%%%%%%%%%%%%%%%%%%%%%%%%%%%%%%%%%%%%%%%%%%%%%%%%%
%%%%%%%%%%%%%%%%%%%%%%%%%%%%%%%%%%%%%%%
%%%%%%%%%%%%%%%%%%%%%%%%%%%%%%%%%%%%%%%
\appendix
%%%%%%%%%%%%%%%%%%%%%%%%%%%%%%%%%%%%%%%
%%%%%%%%%%%%%%%%%%%%%%%%%%%%%%%%%%%%%%%

%%%%%%%%%%%%%%%%%%%%%%%%%%%%%%%%%%%%%%%
\section{Conventions and formalisms}
\label{app:convention}
\subsection*{Massive formalism}
%%%%%%%%%%%%%%%%%%%%%%%%%%%%%%%%%%%%%%%
For a massive particle with mass $m_i$ and momentum $p_i^{\mu}$, the momentum satisfy
%%%%%%%%%%%%%%%%%%%%%%%%%%%%%%%%
\begin{equation}
    \text{det} (p_{i\mu} \sigma_{a\dot{a}}^{\mu})  = \text{det}([p_{i}]_{a\dot{a}}) = m^2_i
    \label{eq:det}
\end{equation}
%%%%%%%%%%%%%%%%%%%%%%%%%%%
The non-zero determinant allows us to represent momentum as
%%%%%%%%%%%%%%%%%%%%%%%%%%%%%%%%%%%%%%%\
%%%%
\begin{align}
	[p_{i}]_{a\dot{a}} = \vert \textbf{i}\rangle_{a}^I [\textbf{i}\vert_{\dot{a}I},
	\quad
	I = 1,2.
    \label{eq:bi-sp}
\end{align}
We use \textbf{bold} notation for massive spinors to distinguish them from massless spinors. $I$ labels the little group $SU(2)$ index, and momentum is invariant under such $SU(2)$ transformations. 

We raise and lower indices using totally anti-symmetric tensors $\varepsilon^{IJ},\varepsilon^{ab},\varepsilon^{\dot{a}\dot{b}}$ satisfying $\varepsilon^{12}=-\varepsilon_{12}=1$. Therefore, we have
\begin{equation}
    \langle \textbf{i} |^a_I=\varepsilon^{ab}\varepsilon_{IJ}|\textbf{i}\rangle^{J}_b,\quad [\textbf{i}|_{\dot a I} = \varepsilon_{\dot a \dot b}\varepsilon_{IJ}|\textbf{i}]^{\dot b J}.
\end{equation}
%%%%
Then, using \eqref{eq:det}, we can impose the relative normalization for the determinants
%%%%%%%%%%%%%%%%%%%
%%
%%
\begin{align}
	\text{det}(\vert \textbf{i}\rangle_{a}^I) = \text{det}([\textbf{i}\vert_{\dot{a}I}) = m_i.
\end{align}
The normalization of the determinant then fixes the following relations among the spinors 
%%%%%
\begin{equation}
    \begin{aligned}
        & |\textbf{i}\rangle_a^I\langle \textbf{i}|^b_I=-m_i\delta_a^b, \quad |\textbf{i}]^{\dot a I}[\textbf{i}|_{\dot b I}=m_i\delta_{\dot b}^{\dot a},\\
        & |\textbf{i}\rangle_a^I[\textbf{i}|_{\dot b I}=p_{ia\dot b}, \quad \quad  |\textbf{i}]^{\dot a I}\langle \textbf{i}|^b_I=-p^{\dot a b}_i.
    \end{aligned}
\end{equation}
%%%%%%%
These definitions further give us the on-shell relations for the spinors
%%%%%%%%%%%
\begin{equation}
    \begin{aligned}
        & p_{ib \dot a}|\textbf{i}]^{\dot a I}=m_i |\textbf{i}\rangle_b^I, \quad p^{\dot b a}_i|\textbf{i}\rangle_a^I=m_i|\textbf{i}]^{\dot b I},\\
        & [\textbf{i}|_{\dot b I}p^{\dot b a}_i=-m_i\langle \textbf{i}|^a_I, \quad \langle \textbf{i}|^b_I p_{ib \dot a}=-m_i [\textbf{i}|_{\dot a I}.
    \end{aligned}
    \label{eq:dirac}
\end{equation}
%%%%%%%%%%%
We can construct Lorentz invariant and little group covariant\footnote{Little group covariance reflects the freedom in the choice of spin axis. } objects by contracting the  \( SL(2, \mathbb{C}) \) indices
%%%%%%%%%
%%
\begin{align}
	\langle \textbf{i}\textbf{j}\rangle^{IJ} 
	= \varepsilon^{ab}\vert \textbf{i}\rangle_{b}^I \vert \textbf{j}\rangle_a^J,
	\quad
	[\textbf{i}\textbf{j}]_{IJ} 
	= \varepsilon_{\dot{a}\dot{b}}\vert \textbf{j}]^{\dot{a}}_{I}\vert \textbf{i}]^{\dot{b}}_J.
\end{align}
%%
%%%%%%%%
Finally, we can enforce proper parity transformation of momentum in \eqref{eq:bi-sp} by imposing the following transformation
%%%%%%
%%
\begin{align}
	\vert -\textbf{i}\rangle^I = \vert \textbf{i}\rangle^I,
	\quad
	\vert -\textbf{i}]_I = -\vert \textbf{i}]_I.
	\label{eq:spinor_negative_mom}
\end{align}
where $\vert -\textbf{i}\rangle^I$ and $\vert -\textbf{i}]_I$ represent spinors associated with $-p_i^{\mu}$, or in other words, they describe the corresponding spinors after time and parity revarsal operation.
%%%%%%%%%
\subsection*{Massless formalism}
%%%%%%%%%%%%%%%%%%%%%%%%%%%%%%%%%%%%%%%
For a massless momentum satisfying \( p^2 = 0 \), we find that  $
\text{det} (p_{i\mu} \sigma_{a\dot{a}}^{\mu}) = 0$.
Then, we can represent the momentum as  
%%%%%%%%%%%%%%%%%%%
%%%%
\begin{align}
	[p_{i}]_{a\dot{a}} = \vert {i}\rangle_{a}[{i}\vert_{\dot{a}}
    \label{eq:bi-sp-massless}
\end{align}
The little group of a massless particle is $U(1)$, which is defined by the freedom to rescale the spinor variables \( \lvert i \rangle \) and \( \lvert i] \) by an overall phase. The angle and bracket spinors transform with opposite helicity weights under the helicity operator. Specifically, we will assign helicity weights of \( +1 \) to \( \lvert i] \) and \( -1 \) to \( \lvert i \rangle \), respectively. 
%%%%%%
\subsection*{Dirac Spinors}
%%%%%%%%%%%%%%%%%%%%%%%%%%%%%%%%%%%%%%%%%%%

Dirac spinors can be written in the form: 
\begin{align}
	u^I(p_i) = \begin{pmatrix} \vert \textbf{i}\rangle_a^{I} \\ \vert \textbf{i}]^{\dot{a}I}\end{pmatrix},&
	\quad
	v^I(p_i) = \begin{pmatrix} \vert \textbf{i}\rangle_a^{I} \\ -\vert \textbf{i}]^{\dot{a}I}\end{pmatrix}, \\
    		\bar{u}^I(p_i) = \begin{pmatrix} -\langle \textbf{i}\vert^{aI} & [\textbf{i}\vert_{\dot{a}}^{I} \end{pmatrix},&
	\quad
	\bar{v}^{I}(p_i) = \begin{pmatrix} \langle \textbf{i}\vert^{aI}  &[\textbf{i}\vert_{\dot{a}}^{I}\end{pmatrix}.
\end{align}
They satisfy the Dirac equation
\begin{align}
	\left[\slashed{p}-m\right] u(p) = 0,&
	\quad
	\left[\slashed{p}+m\right] v(p) = 0,
\end{align}
and completeness relation
\begin{align}
	\varepsilon_{IJ}u^I(p)\bar{u}^J(p) = \slashed{p} + m,
	\quad
	\varepsilon_{IJ}v^{I}(p)\bar{v}^{J}(p) = \slashed{p}-m. \label{eq:fermion_complt}
\end{align}
%%
%%%%%%%%%%
\subsubsection*{Helicity basis and polarization vector}
%%%%%%%%%%
For arbitrary momentum $p$ with solid angle $(\theta,\phi)$, we find the helicity operator
\begin{align}
	\hat{h} = \frac{\hat{p}\cdot \vec{\sigma}}{2}
	= \frac{1}{2}\begin{pmatrix} \cos\theta & \sin\theta e^{-i\phi} \\ \sin\theta e^{i\phi} & -\cos\theta\end{pmatrix}.
\end{align}
The two eigenstates of the operator are found to be:
\begin{align}
	\chi^+ = \begin{pmatrix} c \\ s \end{pmatrix},
	\quad
	\chi^- = \begin{pmatrix} -s^* \\ c \end{pmatrix},
	\quad
	\mathrm{where}
	~~
	c = \cos\frac{\theta}{2},
	~~
	s = e^{i\phi}\sin\frac{\theta}{2},
\end{align}
Using helicity eigenstates, we can represent the bi-spinors in the helicity basis:
%%%
\begin{align}
	&\vert {\mathbf{p}}\rangle_a^{I} = \sqrt{E-|\vec p|}\begin{pmatrix} c \\ s \end{pmatrix} \delta^{I}_+ 
	+ \sqrt{E+|\vec p|}\begin{pmatrix} -s^* \\ c \end{pmatrix} \delta^{I}_-,
	\\
	&[\mathbf{p}\vert_{\dot{a}}^{I} = \sqrt{E+|\vec p|}\begin{pmatrix} -s & c \end{pmatrix} \delta^{I}_+
	- \sqrt{E-|\vec p|}\begin{pmatrix} c & s^* \end{pmatrix} \delta^{I}_-.
\end{align}
Here, we utilize the notation $I=\pm$ to specify little group index and the notation of Kronecker Delta with $\delta_+^+=\delta_-^-=1$. Since these are eigenstates of the helicity operator, the choice of spin axis is made along the momentum direction. 
We note that
\begin{align}
	\varepsilon_{IJ} \vert {\textbf{p}}\rangle_a^{I}[{\textbf{p}}\vert_{\dot{a}}^{J} = p_{a\dot{a}}.
\end{align}
Throughout the paper, we adopt the notation
\begin{align}
	&\vert i\rangle_a = \sqrt{E_i+|\vec{p}_i|}\begin{pmatrix} -s_i^* \\ c_i \end{pmatrix},
	\quad
	[i\vert_{\dot{a}} = \sqrt{E_i + |\vec{p}_i|}\begin{pmatrix} -s_i & c_i \end{pmatrix},
	\\
	&\vert \eta_i \rangle_a = \sqrt{E_i-|\vec{p}_i|}\begin{pmatrix} c_i \\ s_i \end{pmatrix},
	\quad
	[\eta_i\vert_{\dot{a}} = - \sqrt{E_i-|\vec{p}_i|}\begin{pmatrix} c_i & s_i^* \end{pmatrix}.
\end{align}
In the small mass limit, both $\lvert\eta \rangle, [{\eta}\lvert$ scale as $m$, and therefore, $\lvert i \rangle,[i\lvert$ represents the two helicities of the massless particle. In this notation, we find
\begin{align}
	\langle i \eta_i \rangle &= [i \eta_i] = m_i, \\
    	(p_i)_{a\dot{a}} &= \vert i\rangle_a [i\vert_{\dot{a}} - \vert\eta_i\rangle_a [\eta_i\vert_{\dot{a}}.
\end{align}
%%%%%%%%%%%%%
Spin-1 polarization in general basis is defined as:
\begin{align}
	[\epsilon_i^{IJ}]_{a\dot{a}} = \frac{\sqrt{2}}{m_i}\vert {\textbf{i}}\rangle_a^{\{I} [{\textbf{i}}\vert_{\dot{a}}^{J\}}.
\end{align}
In the helicity basis, the polarization is written as
\begin{align}
	\epsilon_i^{(+)} = \sqrt{2}\frac{\vert \eta_i\rangle [ i \vert}{m_i},
	\quad
	\epsilon_i^{(-)} = -\sqrt{2}\frac{\vert i\rangle [ \eta_i \vert}{m_i},
	\quad
	\epsilon_i^{(L)} = \frac{\vert i\rangle [i \vert + \vert \eta_i\rangle [\eta_i \vert}{m_i}.
\end{align}
The signs are chosen so that we have the following normalization
\begin{align}
	\epsilon^{(+)}_i \cdot \epsilon^{(-)}_i = \epsilon^{(L)}_i \cdot \epsilon^{(L)}_i = -1.
\end{align}
%%

%

%%%%%%%%%%%%%%%%%%%%%%%%%%%%%%%%%%%%%%%
\section{Three-point amplitude}
\label{app:threepoint}
%%%%%%%%%%%%%%%%%%%%%%%%%%%%%%%%%%%%%%%

In this appendix, we discuss the on-shell three-point amplitude of a massive spin-3/2 and spin-5/2 particle coupled to a photon and a graviton,
respectively.

%%%%%%%%%%%%%%%%%%%%%%%%%%%%%%%%%%%%%%%
\subsection{Three-point spin-3/2 photon amplitude}
\label{app:threepoint32}
%%%%%%%%%%%%%%%%%%%%%%%%%%%%%%%%%%%%%%%

We begin with the photon case.
In traditional field theory (hybrid of vector and spinor) notation, the spin-3/2 particle is represented by $u^{\mu}_{\lambda}(p)$ and $\bar{v}^{\nu}_{\lambda}(p)$, where $\lambda= \pm3/2, \pm1/2$ indicates the helicities and $p$ is the momentum satisfying $p^2=m^2$. On-shell, they satisfy
%%%%%%%%%%%%%%%%%%%
\begin{align}
    & \Bar{u}(p) \cdot p = \Bar{u}(p) \cdot \gamma = \Bar{u}^{\mu}(p) (\slashed{p} - m) = 0 \ , \\
    & p \cdot u(p) = \gamma \cdot u = ( \slashed{p} - m) u^{\mu}(p) = 0 \ , \\
        & \Bar{v}(p) \cdot p = \Bar{v}(p) \cdot \gamma = \Bar{v}^{\mu}(p) (\slashed{p} + m) = 0 \ ,  \\
    & p \cdot v(p) = \gamma \cdot v(p) = ( \slashed{p} + m) v^{\mu}(p) = 0 \ , 
\end{align}
%%%%%%%%%%%%%%%%%%%
where we suppress the polarization index. 
The general parity-even on-shell three-point interactions up to dimension-five operators are given by
\begin{align}
    \mathcal{A}_3 ( \psi^{3/2}_1, \bar{\psi}^{3/2}_2,  A^{\pm}_3) = \Bar{v}_{\mu}(p_2) \slashed{\epsilon}_3 u^{\mu}(p_1) + \frac{l_1}{m} \Bar{v}_{\mu}(p_2) f_3^{\mu \nu} u_{\nu}(p_1) + \frac{l_2}{m} \Bar{v}_{\mu}(p_2) \tilde{f}_3 u^{\mu}(p_1) \ ,
   \label{eq:spin-3/2 a3}
\end{align}
%%%%%%%%%%%%%%%
where $\epsilon_3$ is the photon polarization, $f_3^{\mu \nu} = p_3^{\mu } \epsilon_3^{\nu} -  p_3^{\nu } \epsilon_3^{\mu}$, 
and $\tilde{f}_3 = f^{\mu \nu}_3 \gamma_{\mu}\gamma_{\nu}$, and we set $m_1=m_2=m$. These terms correspond to the on-shell interactions of the Lagrangian presented in Ref.~\cite{Deser:2000dz}.
Eq.~\eqref{eq:spin-3/2 a3} fulfills the on-shell Ward identity, $\mathcal{A}_3|_{\epsilon_3 \rightarrow p_3} = 0$. The coefficients $l_1$ and $l_2$ can be fixed by the off-shell current constraint\footnote{
%%%%%%%%%%%%%%
Current constraint can ensure factors of $p/m$ in the propagators do not contribute to additional mass divergences, thus ensuring improved high-energy behavior of the amplitude. 
%%%%%%%%%%%%%%%
}
$P \cdot J = \mathcal{O}(m)$, which picks up $l_1=2$ and $l_2 =-1/2$.~\cite{Deser:2000dz,Chiodaroli:2021eug}. 

In the spinor representation, Eq.~\eqref{eq:spin-3/2 a3} takes the form
%%%%%%%
%%%%%%%%%%%%%%%%%%
\begin{align}
   & {A}_3(\psi_1^{3/2},\bar{\psi}_2^{3/2},A^{-}_3) = \frac{\tilde{x}_{12}}{\sqrt{2} m^2}(l_1 [\textbf{1}\textbf{2}]^3 + 2(2l_2 +1)[\textbf{1}\textbf{2}]^2 \langle \textbf{1}\textbf{2}\rangle - (l_1+4l_2)[\textbf{1}\textbf{2}] \langle \textbf{1}\textbf{2}\rangle^2 ) \ \label{A3photoneq1},\\
   & {A}_3(\psi_1^{3/2},\bar{\psi}_2^{3/2},A^{+}_3) = \frac{x_{12}}{\sqrt{2}m^2}(l_1 \langle \textbf{1}\textbf{2}\rangle^3 + 2(2l_2 +1)[\textbf{1}\textbf{2}] \langle \textbf{1}\textbf{2}\rangle^2-(l_1+4l_2)[\textbf{1}\textbf{2}]^2 \langle \textbf{1}\textbf{2}\rangle ) \ .\label{A3photoneq2}
\end{align}
%%%%%%%%%%%%%%%%%%%%%%%%
The most general three-point amplitude also contains $\langle \textbf{1}\textbf{2}\rangle^3$ or $[\textbf{1}\textbf{2}]^3$, 
but these terms arise from operators with dimension higher than 5, which are not considered in this work.
The second term that survives after setting $l_1 = l_2 = 0$ corresponds to the dimension-four operator. 
The current constraint sets $l_1=2$ and $l_2 =-1/2 $ and gives the minimal three-point amplitude as
%%%%%%%%%%%%%%%%%%
\begin{align}
  {A}_3(\psi_1^{3/2},\bar{\psi}_2^{3/2},A^{-}_3)\bigg|_{\mathrm{min.}} = \frac{\sqrt{2}\tilde{x}_{12}}{m^2}[\textbf{1}\textbf{2}]^3 \ ,
   \quad
   {A}(\psi_1^{3/2},\bar{\psi}_2^{3/2},A^{+}_3)\bigg|_{\mathrm{min.}} =  \frac{\sqrt{2}x_{12}}{m^2}\langle \textbf{1}\textbf{2}\rangle^3  \ .
\end{align}
%%%%%%%%%%%%%%%%%%%%%%%%

%%%%%%%%%%%%%%%%%%%%%%%%%%%%%%%%%%%%%%%
\subsection{Three-point spin-5/2 graviton amplitude}
\label{app:threepointspin52}
%%%%%%%%%%%%%%%%%%%%%%%%%%%%%%%%%%%%%%%

Next, we discuss the graviton case.
A spin-5/2 field is represented by a symmetric tensor-spinor  $\psi_{\mu\nu}$, decomposed into a product of polarization vectors and a fermion wavefunction as $\epsilon_\mu \epsilon _\nu u$. An on-shell spin-5/2 particle satisfies
%%%%%%%%
\begin{equation}
    \begin{aligned}
        &\left(\slashed{p}-m\right)\psi_{\mu\nu}=0,%%
        \\&p^\mu \psi_{\mu\nu}=0 ,\quad \gamma^\mu\psi_{\mu\nu}=0,
    \end{aligned}
\end{equation}
which can be obtained from the decomposition of $\psi_{\mu \nu}$ in terms of polarizations and spinor $\epsilon_\mu, \epsilon _\nu, u$. The Lagrangian formulation for a free massive spin-5/2 particle involves the coupling to an auxiliary spin-1/2 particle~\cite{Singh:1974rc}. 
In the curved spacetime formulation, the spin-5/2 particle couples to the graviton through, e.g.~covariant derivatives~\cite{Porrati:1993in}. Little-group covariance prescribes the following spin-5/2-spin-5/2-graviton amplitude:%
\begin{equation}
    \begin{aligned}
        M^{+2}=&\sum _{i=0}^5 g_i x^2_{12} \left<\boldsymbol{1}\boldsymbol{2}\right>^{5-i}\left[\boldsymbol{1}\boldsymbol{2}\right]^i
\\=&
\left(\left<\boldsymbol{1}\epsilon_3^+\boldsymbol{2}\right]+\left<\boldsymbol{2}\epsilon_3^+\boldsymbol{1}\right]\right)^2 
\left(\tilde{g}_0\left<\boldsymbol{1}\boldsymbol{2}\right>^3+\tilde{g}_1 \left<\boldsymbol{1}\boldsymbol{2}\right>^2\left[\boldsymbol{1}\boldsymbol{2}\right] +\tilde{g}_2 \left<\boldsymbol{1}\boldsymbol{2}\right>\left[\boldsymbol{1}\boldsymbol{2}\right] ^2+\tilde{g}_3  \left[\boldsymbol{1}\boldsymbol{2}\right] ^3\right)\\&+ \tilde{g}_4
\left(\left<\boldsymbol{1}\epsilon_3^+\boldsymbol{2}\right]+\left<\boldsymbol{2}\epsilon_3^+\boldsymbol{1}\right]\right)\left[\boldsymbol{1}3\right]\left[\boldsymbol{2}3\right]\left[\boldsymbol{1}\boldsymbol{2}\right]^3+\tilde{g}_5\left[\boldsymbol{1}3\right]^2\left[\boldsymbol{2}3\right]^2\left[\boldsymbol{1}\boldsymbol{2}\right]^3
,   \end{aligned}\label{ampgrav52}
\end{equation}
for positive helicity graviton, with\begin{equation}
    \tilde{g}_0=g_0,\quad \tilde{g}_1=g_1,\quad\tilde{g}_2=g_2, \quad \tilde{g}_3=g_3+g_4+g_5, \quad \tilde{g}_4=\frac{g_4}{m}+\frac{2g_5}{m},\quad \tilde{g}_5=\frac{g_5}{m^2}.
\end{equation}
Minimal coupling corresponds to the first term in Eq.~\eqref{ampgrav52}.  
Notice that interactions with at least two derivatives are necessary to construct the minimal coupling, 
since the polarization vector of the spin-5/2 particle contains either 3 $\left|\boldsymbol{1}\right>$'s and 2 $\left|\boldsymbol{1}\right]$'s, or 2 $\left|\boldsymbol{1}\right>$'s and 3 $\left|\boldsymbol{1}\right]$'s, 
and in the $\tilde{g}_0$ term of Eq.~\eqref{ampgrav52}, $\left|\boldsymbol{1}\right>$ can appear 5 times, which necessitates at least 2 derivatives to convert ``square'' into   ``angle'', upon using the Dirac equation in Eq.~\eqref{eq:dirac}.  With similar arguments, we can conclude that the $\tilde{g}_4$ and $\tilde{g}_5$ amplitudes in Eq.~\eqref{ampgrav52} originate from interactions with more than two derivatives.

The appearance of two derivative operators may seem odd at first sight. The spin-5/2 Lagrangian involves at most one derivative in the kinetic term, so the usual linear expansion of the metric yields the coupling $h_{\mu\nu}T^{\mu\nu}$ which also has up to one derivative. In fact, the necessity of two-derivative couplings has  already been pointed out in \cite{Porrati:1993in}, which showed that  
 $h_{\mu\nu}T^{\mu\nu}$ violates unitarity in the spin-5/2 gravitational scattering amplitudes, and to satisfy the current constraint (for better high energy behavior), one needs to introduce additional couplings to Riemann tensor, e.g., $R_{\mu\nu \alpha\beta}\bar{\psi}^{\mu\alpha}\psi^{\nu\beta }$. Those terms contain two derivatives. In terms of Lagrangian interactions, the two derivative couplings can be:
\begin{equation}
\begin{aligned}&
h_{\mu\nu}\partial_\alpha\bar{\psi}^{\beta \mu }\left(   \rho_1 +i \rho_2 
 \gamma^5    \right)\partial_\beta  \psi^{\nu\alpha }+ h_{\mu\nu} \partial^\nu \bar{\psi}_{\alpha\beta}\left(\rho_3 +i \rho_4\gamma^5\right)\partial^\mu\psi^{\alpha\beta} \\&+h_{\mu\nu } \partial_\beta \bar{\psi}_\alpha^\nu \left(\rho_5+i \rho_6\gamma^5 \right) \partial^\mu \psi^{\beta\alpha}+h_{\mu\nu }\partial^\mu \bar{ \psi}^{\beta\alpha}\left(\rho_5+i\rho_6\gamma^5 \right) \partial_\beta  {\psi}_\alpha^\nu\\&+h_{\mu\nu }  \bar{\psi}_{\alpha \beta} \left(\rho_7+i\rho_8\gamma^5 \right) \partial^\alpha \partial^\beta\psi^{\mu\nu}+h_{\mu\nu }  \partial^\alpha \partial^\beta \bar{\psi}^{\mu \nu} \left(\rho_7+i\rho_8\gamma^5 \right)\psi_{\alpha\beta},
\end{aligned}
\end{equation}
where $\rho$s are real-valued parameters.

\bibliographystyle{utphys}
\bibliography{references}
%%%%%%%%%%%%%%%%%%%%%%%%%%%%%%%%%%%%%%%
  
%%%%%%%%%%%%%%%%%%%%%%%%%%%%%%%%%%%%%%%
\end{document}